\documentclass[twocolumn,trackchanges]{aastex63}

\usepackage{soul}
\usepackage[utf8]{inputenc}
\usepackage{amsmath}
\usepackage{multirow}
\usepackage{threeparttable}
\usepackage{verbatim}
\usepackage{color}

\def\spose#1{\hbox to 0pt{#1\hss}}
\newcommand\lsim{\mathrel{\spose{\lower 3.0pt0pt\hbox{$\mathchar"218$}}
     \raise 2.0pt\hbox{$\mathchar"13C$}}}
\newcommand\gsim{\mathrel{\spose{\lower 3.0pt\hbox{$\mathchar"218$}}
     \raise 2.0pt\hbox{$\mathchar"13E$}}}
\newcommand\msun{{\rm \,M_\odot}}

\newcommand\Msun{{\mathinner{~ \mathrm{M_\odot}}}}

\newcommand\cMpc{{\mathinner{\mathrm{cMpc}}}}
\newcommand\kpc{{\mathinner{\mathrm{kpc}}}}

\definecolor{green}{rgb}{0.0, 0.4, 0.0}
\definecolor{forestgreen(web)}{rgb}{0.13, 0.55, 0.13}
\definecolor{green(web)}{rgb}{0.13, 0.55, 0.13}
\definecolor{green}{rgb}{0.0, 0.4, 0.0}

\def\apjl{Astrophys.\ J.\ Lett.}
\def\mnras{Mon.\ Not.\ R.\ Astron.\ Soc.}

\def\araa{Ann. Rev. Astron. Astrophys.}
\def\aap{Astron.\ Astrophys.}
\def\apj{Astrophys.\ J.}

\def\apjs{Astrophys.\ J. Supp.}

\def\prd{Phys.\ Rev.\ D}

\def\jcap{{JCAP\ }}

\def\pasj{PASJ}

\shorttitle{Horizon Run 5}
\shortauthors{Lee \& Shin et~al.}

\begin{document}
\title{The Horizon Run 5 Cosmological Hydrodynamical Simulation:  \\Probing Galaxy Formation from Kilo- to Giga-parsec Scales}

\correspondingauthor{Jihye Shin}
\email{jhshin@kasi.re.kr}

\author[0000-0002-6810-1778]{Jaehyun Lee}
\altaffiliation{Jaehyun Lee and Jihye Shin contributed equally to this work as first authors.}
\affiliation{School of Physics, Korea Institute for Advanced Study (KIAS), 85 Hoegiro, Dongdaemun-gu, Seoul 02455, Republic of Korea}

\author[0000-0001-5135-1693]{Jihye Shin}
\altaffiliation{Jaehyun Lee and Jihye Shin contributed equally to this work as first authors.}
\affiliation{Korea Astronomy and Space Science Institute (KASI), 776 Daedeokdae-ro, Yuseong-gu, Daejeon 34055, Republic of Korea}

\author{Owain N. Snaith}
\affiliation{GEPI, Observatoire de Paris, PSL Universit\'e, CNRS, 5 Place Jules Janssen, 92190, Meudon, France}

\author{Yonghwi Kim}
\affiliation{School of Physics, Korea Institute for Advanced Study (KIAS), 85 Hoegiro, Dongdaemun-gu, Seoul 02455, Republic of Korea}

\author{C. Gareth Few}
\affiliation{E.A. Milne Centre for Astrophysics, University of Hull, Hull, HU17 7HB, United Kingdom}
\affiliation{Department of Physics, Durham University, South Road, Durham DH1 3LE, United Kingdom}

\author[0000-0002-8140-0422]{Julien Devriendt}
\affiliation{Department of Physics, University of Oxford, Keble Road, Oxford, OX1 3RH, United Kingdom}

\author[0000-0003-0225-6387]{Yohan Dubois}
\affiliation{CNRS and Sorbonne Universit\'e, UMR 7095, Institut d'Astrophysique de Paris, 98 bis Boulevard Arago, F-75014 Paris, France}

\author[0000-0003-1165-001X]{Leah M. Cox}
\affiliation{E.A. Milne Centre for Astrophysics, University of Hull, Hull, HU17 7HB, United Kingdom}

\author[0000-0003-4923-8485]{Sungwook E. Hong}
\affiliation{Korea Astronomy and Space Science Institute (KASI), 776 Daedeokdae-ro, Yuseong-gu, Daejeon 34055, Republic of Korea}
\affiliation{Natural Science Research Institute, University of Seoul, 163 Seoulsiripdaero, Dongdaemun-gu, Seoul 02504, Korea}

\author{Oh-Kyoung Kwon}
\affiliation{Korea Institute of Science and Technology Information (KISTI), 245 Daehak-ro, Yuseong-gu, Daejeon 34141, Republic of Korea}

\author[0000-0002-2692-7520]{Chan Park}
\affiliation{National Institute for Mathematical Sciences (NIMS), 70 Yuseong-daero 1689 beon-gil, Yuseong-gu, Daejeon 34047, Republic of Korea}

\author{Christophe Pichon}
\affiliation{School of Physics, Korea Institute for Advanced Study (KIAS), 85 Hoegiro, Dongdaemun-gu, Seoul 02455, Republic of Korea}
\affiliation{CNRS and Sorbonne Universit\'e, UMR 7095, Institut d'Astrophysique de Paris, 98 bis Boulevard Arago, F-75014 Paris, France}

\author[0000-0002-4391-2275]{Juhan Kim}
\affiliation{Center for Advanced Computation, Korea Institute for Advanced Study, 85 Hoegiro, Dongdaemun-gu, Seoul 02455, Republic of Korea}

\author[0000-0003-4446-3130]{Brad K. Gibson}
\affiliation{E.A. Milne Centre for Astrophysics, University of Hull, Hull, HU17 7HB, United Kingdom}

\author[0000-0001-9521-6397]{Changbom Park}
\affiliation{School of Physics, Korea Institute for Advanced Study (KIAS), 85 Hoegiro, Dongdaemun-gu, Seoul 02455, Republic of Korea}

\begin{abstract}
Horizon Run 5 (\texttt{HR5}) is a cosmological hydrodynamical simulation which captures the properties of the Universe on a Gpc scale while achieving a resolution of 1~kpc. 
Inside the simulation box we zoom-in on a high-resolution cuboid region with a volume of $1049\times119\times127 \, {\rm cMpc^3}$.
The sub-grid physics chosen to model galaxy formation includes radiative heating/cooling, UV background, star formation, supernova feedback, chemical evolution tracking the enrichment of oxygen and iron, the growth of supermassive black holes and feedback from active galactic nuclei (AGN) in the form of a dual jet-heating mode.  For this simulation we implemented a hybrid \texttt{MPI-OpenMP} version of \texttt{RAMSES}, specifically targeted for modern many-core many thread parallel architectures. In addition to the traditional simulation snapshots, light-cone data was generated on the fly. For the post-processing, we extended the Friends-of-Friend (FoF) algorithm and developed a new galaxy finder  \texttt{PGalF} to analyse the outputs of \texttt{HR5}.
The simulation successfully reproduces observations, such as the cosmic star formation history and connectivity of galaxy distribution, We identify cosmological structures at a wide range of scales, from filaments with a length of several $\cMpc$, to voids with a radius of $\sim100\,\cMpc$.
The simulation also indicates that hydrodynamical effects on small scales impact galaxy clustering up to very large scales near and beyond the baryonic acoustic oscillation (BAO) scale. Hence, caution should be taken when using that scale as a cosmic standard ruler: one needs to carefully understand the corresponding biases. The simulation is expected to be an invaluable asset for the interpretation of upcoming deep surveys of the Universe. 
\end{abstract}


\section{Introduction} \label{sec:intro}
Understanding the cosmic origin of the observed diversity of galaxies is an interesting and challenging problem.
While baryonic matter accounts for only about 5\% of the energy budget of the universe, its impact on galaxy formation is critical, and must be accounted for  statistically.
Over the last decade a plethora of  cosmological simulations reaching typically kiloparsec resolution  have been produced to address this challenge (\texttt{MareNostrum} \citep{Ocvirk2008}, \texttt{Horizon-AGN} \citep{Dubois2014}, \texttt{Illustris} \citep{Genel2014}, \texttt{MassiveBlack-II} \citep{Khandai15}, \texttt{EAGLE} \citep{Schaye2015}, \texttt{Magneticum} \citep{Dolag2016}, \texttt{Romulus} \citep{Tremmel2017MNRAS},  \texttt{BAHAMAS}  \citep{bahamas2017}, \texttt{IllustrisTNG} \citep{Springel18,Pillepich2018,Naiman18,Marinacci18,Nelson18,Nelson19}, \texttt{SIMBA} \citep{Dave2019}).
They aim to model as accurately as possible the intricate physical processes occurring on multiple scales, either by resolving them (using refinement techniques) or  using so-called sub-grid models. They track the full cosmic history of what aims to be a statistically  representative region of the Universe  using hydrodynamics and gravity and typically account for gas cooling,  star formation,  stellar and AGN feedback as well as metal production, in order to provide statistical insights into a wide range of astrophysical problems. 

We are also fortunate to live in the age of existing and forthcoming  very large photometric (SXDS \citep{SXDS2004}, COSMOS \citep{Scoville2007}, Alhambra \citep{ALHAMBRA2005}, DES \citep{DES2014},  Euclid \citep{Euclid}, HSC \citep{HSC2018...70S...4A} and LSST \citep{LSST}) and spectroscopic surveys (4MOST \citep{4MOST},  JPAS \citep{JPAS2014}, WFIRST \citep{Spergel2013}, DESI \citep{DESI}, MSE \citep{MSE2016}, PFS \citep{Aihara2018})\footnote{
https://www.desi.lbl.gov,\\
https://www.euclid-ec.org,\\
https://www.lsst.org,\\ https://pfs.ipmu.jp,\\
https://wfirst.gsfc.nasa.gov}, allowing us to probe not only our present day Universe, but also a significant period of cosmic time.

The confrontation of the cosmological simulations with galaxy surveys has been very successful at reproducing a significant number of features. Examples include studies of the amplitude of galaxy clustering, the morphology and topology of large-scale structures \citep{1990MNRAS.242P..59P, 2005ApJ...633...11P, 2012ApJ...759L...7P,2010ApJS..190..181C}, the cosmic evolution of the star formation rate and luminosity function \citep[e.g.][]{Devriendt2010}, the bimodality of the physical, spectroscopic and morphological properties of galaxies at low redshift \citep[e.g.][]{Dubois2016}, while probing a diversity of  environments and epochs.  Nonetheless, compared to the observed Universe, one of the main limitations of past cosmological hydrodynamical simulations is the dynamical range of scales probed. 

Recent examples of simulations of entire cosmological volumes, such as  \texttt{Horizon-AGN}, the \texttt{TNG100} run of the \texttt{IllustrisTNG} project or \texttt{Eagle}, can capture scales ranging from $\sim 100\,{\rm Mpc}$ to $1\,{\rm kpc}$.  This is rather restrictive, both from a statistical standpoint --- rare events are quite sensitive to the underlying cosmological parameters --- but also from a physical standpoint. In particular, the large-scale peculiar velocity field cannot be properly recovered for many cosmological models, such as the popular $\Lambda$CDM model, in such small-volume simulations. Furthermore, the gravity and baryonic processes couple over widely different scales (as can be seen in intrinsic alignments, or in the strangulation of dwarfs in clusters). When improperly accounted for this can, in turn, impact dark energy experiments \citep[e.g.][]{Chisari2018}, among others. Indeed, one significant shortcoming of most past simulations is to underestimate the  observed spread in the properties of cosmic structures  (e.g. colours, $V/\sigma$ etc.), which could be due to simulators calibrating their sub-grid physics on the mean of the observed process (and not accounting for its full variance), but could also be a consequence of the lack of diversity of the underlying physics captured in small boxes (such as the lack of rare events). For instance, the mean separation of rich clusters is known to be $\sim70\,\cMpc$ and $\sim40\,\cMpc$ for QSOs \citep{bahcall92}, indicating that only a few dense environments and QSOs are contained in a cosmological volume with a side length of 100 cMpc. Furthermore, \citet{colombi94} show that a two-point correlation function begins to deviate from theory when the separation scale is larger than one fifth of simulation box size. Therefore, one may need simulations with a volume larger than hundreds of cMpc on the side in order to examine the distribution of galaxies on the BAO scale, and to capture a wide range of environments.

The Horizon Run series of cosmological simulations \citep{2009ApJ...701.1547K,2011JKAS...44..217K,2015JKAS...48..213K} have adopted large volumes ($84-3390\,{\rm cGpc^3}$) in order to grasp the important large-scale properties of the cosmological models under study, as well as to secure large statistical samples for quantitative comparisons between observations and simulations. The Horizon Run 5 (\texttt{HR5}) simulation presented in this paper, being the first hydrodynamic simulation in this series, follows the idea of the previous generations of the Horizon Runs.  \texttt{HR5} is able to capture physical processes on scales ranging from Gpc down to kpc scales, with a dynamic range an order of magnitude greater than any previous simulations of this kind.
The total volume of the high resolution region is $\sim$2 times smaller than the largest \texttt{TNG300} simulation of the \texttt{IllustrisTNG} project 
\citep{Springel18} 
with slightly better resolution. However, the true strength of \texttt{HR5} is that the length of the simulation box will give us access to the long wavelength modes of the density power spectrum up to $1\,{\rm cGpc}$. This is something that has not previously been achievable in hydrodynamic simulations that are also capable of resolving galactic processes.

 As discussed in more detail below, the geometry of the simulation was chosen to provide means of easily extracting  virtual lightcones from the data. These can easily be used for Ly-$\alpha$ tomography, and more generally to study the large-scale chemical  properties of the IGM. Compared to other hydrodynamical simulations, the wide range of scales probed by \texttt{HR5} provides us with a large-scale peculiar velocity field consistent with the underlying cosmological model, a fair sample of massive clusters, and allows us to probe the impact of the very large-scale structures (wall, filaments, voids) on galaxy formation,  clustering and weak lensing. In particular, the simulation was customized to probe BAO scales, the impact of large-scale streaming velocity, velocity bias, and intrinsic alignments. This is achieved by setting up initial conditions which reflect the  expected excess power at the relevant scale. The simulation's cosmic variance is also set to obey that of the Universe on Giga-parsec scales, which allows us to study the formation of massive clusters and 
the most massive SMBH they host in a realistic environment. We implemented a bipolar jet model for AGN feedback coupled to black hole spin, which aims to better model the energy deposition on kpc scales. We also included detailed chemistry (including oxygen and iron) for the IGM  to answer astrophysical problems, such as the missing baryon problem.  At a technical level, we made use of a power spectrum generated by \texttt{CAMB}  \citep{CAP...04..027H} to generate distinct velocity fields for the dark matter  and  baryons, and implemented  \texttt{OpenMP} parallelization in order to make better use of modern multi-core computing architectures.
 
In the following paper we will present the main characteristics of the simulation in 
Sect.~\ref{sec:simulation}, while 
Sect.~\ref{sec:subgrid} describes the subgrid physics prescriptions incorporated into the simulation code.
Sect.~\ref{sec:output} presents the type of raw data produced in the simulation suite.
Sect.~\ref{sec:finding}  presents the suite's dark matter halos, and large-scale structure properties.
In Sect.~\ref{sec:global} we present our results on the the global stellar properties of the simulation. 
Finally, we will present our conclusions in Sect.~\ref{sec:conclusion}.

\section{Simulations} \label{sec:simulation}
We will first describe the numerical code 
used for \texttt{HR5} (Sect.~\ref{sub:numerical}) before focusing on the generation of the zoomed initial conditions (Sect.~\ref{sub:IC}).
We will then discuss the \texttt{OpenMP} optimisation (Sect.~\ref{sub:optim}) and the output strategy.
Finally, sister simulations with slightly different subgrid physics  will be described in Sect.~\ref{sub:suite}.

\subsection{Numerical setup}\label{sub:numerical}

The simulations were carried out using the adaptive mesh refinement code \texttt{RAMSES}~\citep{Teyssier02}.
To solve for gravity, the dark matter, star and black hole (BH) particles masses are projected onto the grid with a cloud-in-cell assignment scheme. The grid mass density (together with the contribution of the gas) is used to compute the gravitational acceleration through the Poisson equation solved with a multi-grid relaxation method. 
Particles are evolved through time using leapfrog integration.
The grid is adaptively refined in the zoom-in region using a quasi-Lagrangian criterion: wherever the mass is larger (smaller) than eight times that of the high-resolution dark matter/baryonic resolution, the cell is refined (or derefined) up to a best-achieved resolution of 1~physical kpc.
To achieve a nearly constant physical maximum resolution in a comoving box, a new level of refinement is added at expansion factors $a_{\rm exp}=0.0125$, 0.025, 0.05, 0.1, 0.2, 0.4 and $0.8$, where the present epoch is defined by $a_{\rm exp}=1$.
The time step is adaptive, with a shared step size across a given refinement level, and varies by a factor of 2 across contiguous levels. The minimum step size is determined by the Courant condition, with a Courant factor of $C=0.8$.

The hydrodynamics is solved directly on the grid, rather than by evolving particles whose potential and acceleration are computed using the grid. The set of Euler equations is solved with the unsplit MUSCL-Hancock method~\citep{vanleer79}: fluxes are obtained with a second-order Godunov scheme using the approximate Harten-Lax-van Leer-Contact~\citep{Toro94} Riemann solver, and the minmod slope limiter on conservative variables. 
The gas is assumed to be of primordial composition, with a hydrogen fraction of $X_{\rm H}=0.76$. The remaining gas is assumed to be composed of helium, and the mixture follows the ideal equation of state 
of a mono-atomic gas with adiabatic index of $\gamma=5/3$.
\begin{figure}
    \centering
    \includegraphics[width=1\columnwidth]{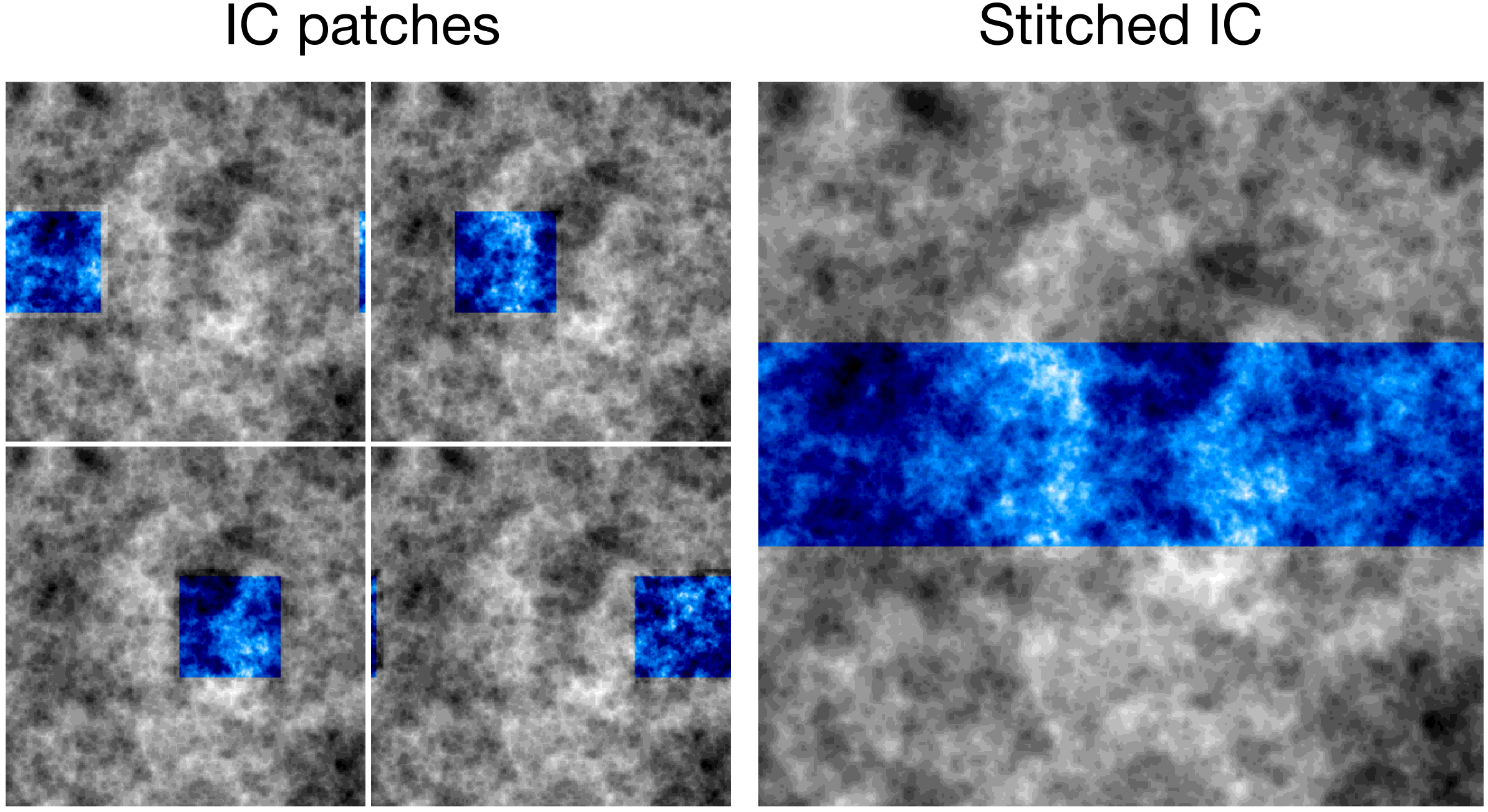} \\
    \caption{Generation of the initial conditions in a cuboid shaped zoomed-in region using the \texttt{MUSIC} package. The blue colour scaling shows the high resolution region of the ICs, while the grey colour scale shows the regions that are not refined. The four left-hand plots show different individual realisations of the high resolution ICs, while the right-hand plots shows the combined final ICs used in the simulation.}\label{fig:ic}
\end{figure}

\begin{figure*}
    \centering
    \includegraphics[width=2\columnwidth]{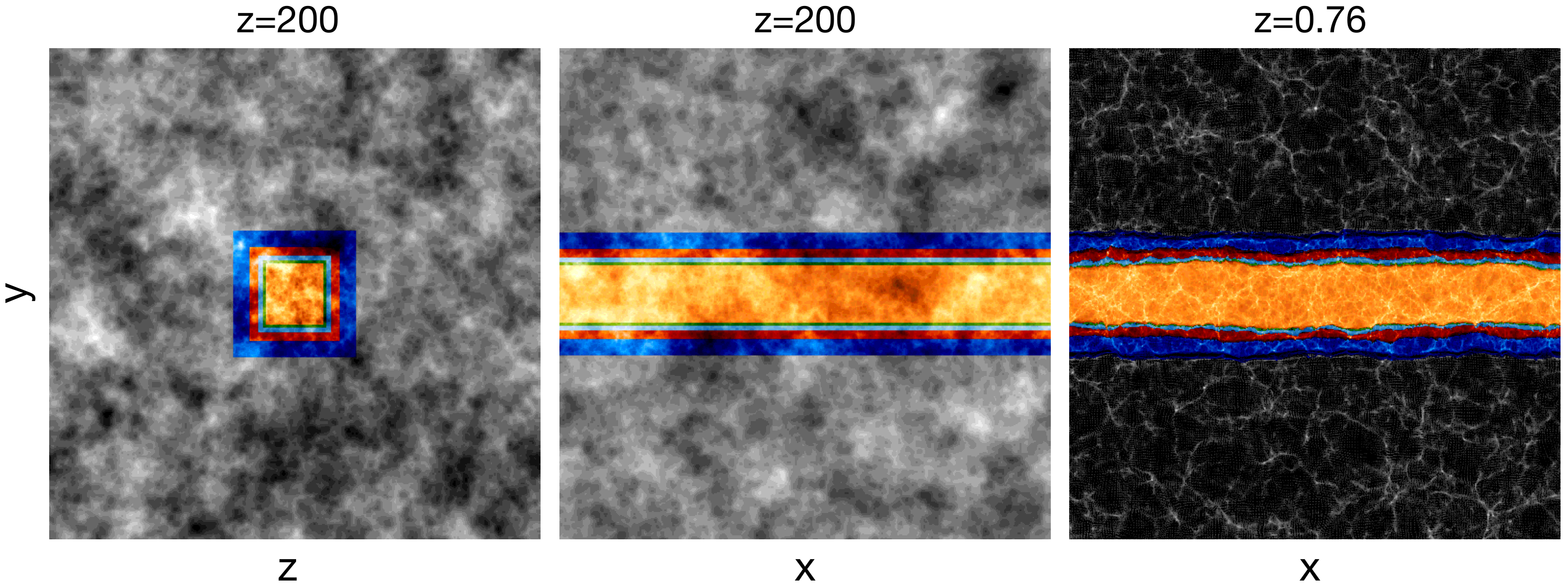} \\
    \caption{Configuration of the zoomed-in region of \texttt{HR5}. The left and middle panels show the zoom region projected onto the $y$--$z$ and $y$--$x$ planes with the size of $1049\,\cMpc$, respectively, at the initial redshift. In the right panel, we present the evolved density at $z=0.76$ to show the large-scale perturbations running across the zoomed region. Different colours illustrate the different zoom levels going from blue to orange, and increase in resolution by a factor of two each time. }\label{fig:zoom}
\end{figure*}

The simulation probes the large-scale structures residing in a cubic volume of $(1049\,\cMpc)^3$ with the highest resolution elements measuring 1~kpc in size. In order to take advantage both of the high-resolution and to sample the very large-scale structures, we defined an elongated cuboidal zoom region whose length is $1049\,\cMpc$ 
and rectangle cross-section measures $119\times127\,\cMpc^2$. 
Outside the zoom region, the low-resolution elements  account for the influence from  the large-scale structure. This unique zoom-in geometry was adopted to optimize  the construction of a light-cone, to facilitate a direct comparison with existing and upcoming surveys of the deep Universe, e.g. COSMOS \citep{Scoville2007}, DES \citep{DES2014}, LSST \citep{LSST}, and DESI \citep{DESI}.

\subsection{Cosmology \& Initial conditions}\label{sub:IC}

The cosmological parameters are compatible with Planck data \citep{Planck2016} with $\Omega_{\rm m}$=0.3, $\Omega_{\Lambda}$=0.7, $\Omega_{\rm b}$=0.047, $\sigma_8=0.816$, and $h_0$=0.684, and the linear power spectrum is calculated from the \texttt{CAMB} package \citep{lew00}. The simulation initial conditions at $z=200$ are generated with the \texttt{MUSIC} package \citep{hah11} using second-order Lagrangian perturbation theory (2LPT; \citealt{sco98,lhu14}).

The initial density field in a periodic box of $1049~\cMpc$ size is generated on a $256^3$ grid where each cell size is $4.09~\cMpc$, while that for the high-resolution zoom region of $1049\times119\times127\,\cMpc^3$ is filled with $8192\times896\times896$ cells with a side length of 128\,ckpc. Inside the zoom region, 128\,ckpc sized cells at ${\rm level} = 13$ are gradually refined up to $1\,\kpc$  at ${\rm level} = 20$ at $z = 0$. Note that the \texttt{HR5} simulation stops at $z = 0.625$, and so the final resolution inside the zoom region is stopped at 2\,ckpc . To avoid low-resolution particles from outside the region of interest contaminating the high-resolution region, four intermediate buffer regions of ${\rm level}=9-12$ surround the zoom region (see Figure~\ref{fig:ic}). Furthermore, \texttt{RAMSES} forbids a jump of more than one level of refinement in contiguous regions. If we wish to increase the resolution by a factor of four, for example, we must produce an intermediate refinement level in between the high and low resolution regions. 

\texttt{MUSIC} internally doubles the size of the region in every dimension, to deal with isolated boundary conditions. To overcome this issue we generated a series of smaller zoom regions with regular offsets under the same realization (see Figure~\ref{fig:ic}), and then stitched them together in order to generate the elongated zoom region. 
More specifically, we generated 16 different zoom regions with a $65.54\,\cMpc$ offset between them along the $x$-direction, and then stitched them to make a single cuboid shaped region, as shown in Figure \ref{fig:zoom}. We address the validity of the stitching scheme and the consistency of the peculiar geometry of the zoomed region in Appendices~\ref{sec:consistency} and~\ref{sec:cosvar}.

\subsection{Performance \& Optimization}\label{sub:optim}

A substantial speedup of the simulation code has been achieved by implementing an additional dimension of parallelism, using \texttt{OpenMP}, on top of the original Message-Passing-Interface (\texttt{MPI}) layer of \texttt{RAMSES}, to fully take advantage of the shared memory architecture and multiple computing threads of single computing nodes.

The \texttt{OpenMP} is orthogonal to \texttt{MPI} in the parallel domain, meaning that there is no interference between the two methods as long as the routines are thread safe.
For this purpose, we deleted the \texttt{save} keyword in the original source code, whenever possible, and made all the local variables thread safe, even though it may have a mild detrimental effect on the performance of the code. We also exchanged the current sequential \texttt{BH merging} routine with the tree searching method, which is designed to be both thread safe and parallel.

We also made considerable changes to the sequential routines by removing \texttt{do-loop}s, unless the code had unavoidable \texttt{atomic} functions. 

Furthermore, we suppressed the use of stack memory, by using the heap memory with dynamic allocation. This is because the \texttt{HR5} simulation uses a large amount of memory, which can lead to stack overflows. These can have unpredictable consequences, leading to crashes during the run. 

\begin{figure}
    \centering
    \includegraphics[width=0.9\columnwidth]{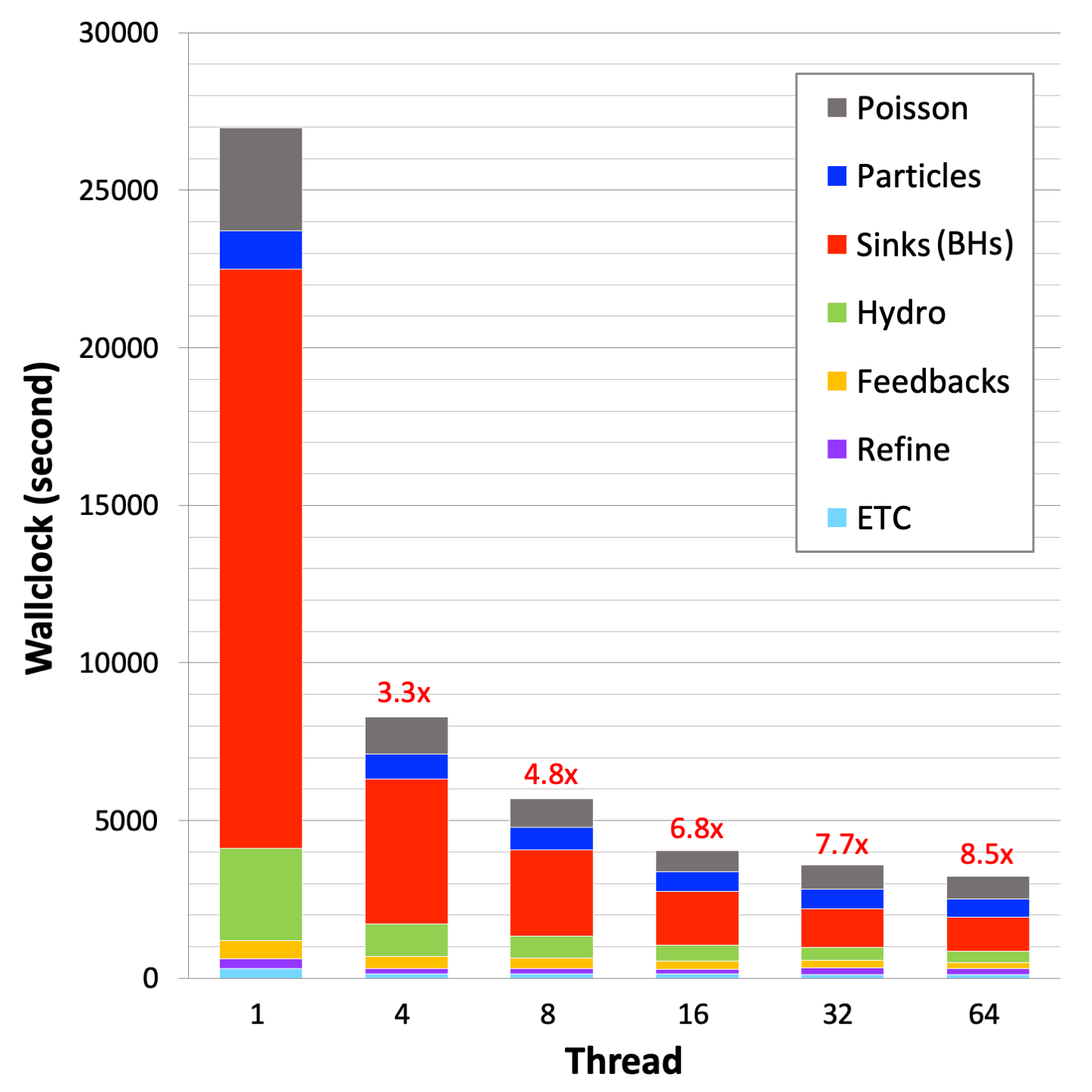}
    \caption{Parallel performance of the code after turning on \texttt{OpenMP}.
    In the legend, the subroutine names are listed in the same order as top of the bar graph. The overall value of the clock-time speed up of the code is included on top of each bar.
    \label{Fig:speedup}
    }
\end{figure}

The overall impact of the number of threads on  run time for the different components of the code are shown in Figure~\ref{Fig:speedup}, which demonstrates improvement with the number of threads, although with diminishing returns. For example, the run with 64~threads consumes  8.5 times less wall-clock time  than the single thread version, but gives us access to 64 times more memory. 

\subsection{A Suite of Simulations }
\label{sub:suite}
To explore the dependence of the simulation results on our chosen cosmological and astrophysical parameters, we have run several companion simulations, with different choices of kinetic SN feedback and delayed cooling (as listed in Table~\ref{tab:hr5}). \texttt{HR5} represents our standard (fiducial) model. The suffix \texttt{DC} indicates the simulation which includes delayed cooling. \texttt{HR5-lowQSO} is a model without delayed cooling, and with a QSO mode efficiency $\epsilon_{\rm f,h}$ $1/10$ of the standard run. These runs were carried out to examine the impact of the QSO mode efficiency on BH growth with spins. One can find the further details of this physical ingredients in Sect.~\ref{sub:agn}. 
\begin{deluxetable}{lccc}[hbt]
\tablehead{
\colhead{Name} &
\colhead{Final $z$} &
\colhead{Delayed Cooling} &
\colhead{$\epsilon_{\rm f,h}$}
}
\label{tab:hr5}
\caption{Parameters of the suite of \texttt{HR5} simulations. $\epsilon_{\rm f,h}$ is the coupling efficiency of thermal (QSO) mode AGN feedback. In addition $M^\textrm{BH}_\textrm{seed}$= $10^4\,\Msun$, $M_{\rm DM} = 6.9\times 10^7\,\Msun$ for the finest DM particles, and the maximal spatial resolution $\sim 1\,\kpc$. The chemical enrichment is traced for elements H, O, and Fe. }
\startdata
\texttt{HR5 (fiducial)} & 0.625 & No  &  0.15 \\
\texttt{HR5-lowQSO} & 1.5 & No   & 0.015 \\
\texttt{HR5-DC} & 3.0 & Yes   & 0.15 
\enddata
\end{deluxetable}

All the runs have a spatial resolution down to $\sim1$ physical kpc. \texttt{HR5} has reached and has been stopped at redshift $z=0.625$, and it is suitable for comparisons to many deep surveys that are currently underway or planned in the near future reaching the universe beyond this redshift. \texttt{HR5} should be useful in designing future surveys and understanding the survey results. Since this paper focuses on announcing \texttt{HR5} simulations, in-depth comparison between the three different runs will be made in follow-up studies.

We performed the suite of \texttt{HR5} simulations using the {\tt Nurion} supercomputer at the Korea Institute of Science and Technology Institute (KISTI), which consists of 8,305~compute nodes, has 797.3~TB of memory, and a storage capacity of 21~PB. Nurion has a theoretical maximum performance of 25.7 Pflops, based on the Intel Xeon Phi many-core processors. Each processor  contains 68 physical cores, and 100 Gbps high-performance interconnections. We had exclusive use of the 2,500~compute nodes for three months, from December 2018 to February 2019. For the first month, we ran \texttt{HR5} using 1,250~compute nodes (2,500~MPI ranks and 32~threads), while utilizing another 1,250~compute nodes to perform \texttt{HR5-lowQSO} and \texttt{HR5-DC}. Since \texttt{HR5} ultimately required more than 120~TB of memory (which exceeds the total memory available on the assigned 1,250~compute nodes) we suspended \texttt{HR5-lowQSO} and \texttt{HR5-DC} and assigned all 2,500~compute nodes to the \texttt{HR5} simulation (and made use of 2,500~MPI ranks and 64~threads) for the following two months. The total data size for the three runs is approximately 2~PB, one-tenth of total capacity of the {\tt Nurion} storage system. 

\section{Subgrid physics}\label{sec:subgrid}
In the following section we present the sub-grid physics implemented in the \texttt{HR5} to model gas cooling and heating (Sect.~\ref{sub:cooling}), AGN (Sect.~\ref{sub:agn}), star formation (Sect.~\ref{sub:formation}),  chemistry (Sect.~\ref{sub:chemistry}), and SN feedback (Sect.~\ref{sub:feedback}). 

In order to set the sub-grid physics parameters we compared the results of the simulations to observations. This is essential, because the results of the simulations are sensitive to these choices, and the best values of parameters vary with implementation method and resolution. In order to find a reasonable parameter set, we made a number of smaller volume simulations, with the same cosmology and resolution as the final simulations. In particular, we chiefly required the cosmic star formation history (CSFH) to match the observational data collected in multiple studies, e.g. \citet{Hopkins04}, \citet{Behroozi13}, \citet{Madau_Dickinson14} (see Figure~\ref{fig:sfh} for a comparison with \texttt{HR5} and more details in Appendix~\ref{sec:simparamtuning}). In practice, the comparison between the test simulations and observations for the parameter evaluation was made by eye, with no formal fitting procedure. Otherwise, computing costs to run a variety of simulations required to make a precise fitting to observations would explode. Besides, this helps us to avoid overfitting, or biasing too much toward our choice of empirical data.

\subsection{Gas cooling and heating}\label{sub:cooling}

The internal energy loss of the primordial and metal-enriched gas is computed down to $\sim 10^4\,$K using the synthetic cooling functions derived by \citet{sutherland93}, based on Bremstrahlung, collisional ionization, recombination, line radiation, and Compton cooling processes. Metal-enriched gas can cool even further below $\sim 10^4\,$K using the cooling rates proposed by \citet{dalgarno72}. In \texttt{HR5}, gas is allowed to cool down to 750\,K. Chemical enrichment is thus traced in a self-consistent manner for better estimating gas cooling rates. The details of the chemical evolution is described in Sect. \ref{sub:chemistry}. A uniform UV background is assumed to approximate reionization, following \citet{haardt96}. This process gradually heats gas in the entire volume after redshift $z_{\rm reion}=10$.

\subsection{AGN feedback}\label{sub:agn}

Black holes (BH) are seeded with an initial mass of $10^4\,\msun$ in regions with gas density above $n_{\rm H,0}=0.1\, {\rm H\, cm^{-3}}$. However, BHs are only produced if there is no other BH within $50\,\kpc$. The dynamics of the BH is corrected for an explicit unresolved gaseous drag term~\citep{Dubois13drag}, $F_{\rm drag}=f_{\rm gas}4\pi \alpha \bar \rho (G M_{\rm BH}/\bar c_{\rm s})^2$, where $\bar \rho$ is the average gas density, $f_{\rm gas}$ is a Mach-number ($\mathcal{M}=\bar u/\bar c_{\rm s}$) dependent factor \citep{Ostriker99}, $\bar u$ and $\bar c_{\rm s}$ are the average BH-gas relative velocity and gas sound speed respectively, $G$ is the gravitational constant and $M_{\rm BH}$ is the BH mass. The drag force experienced by the BH is boosted by $\alpha=(\rho/\rho_0)^2$, in regions where the gas density exceeds the threshold of star formation $\rho_0=n_{\rm H,0}m_{\rm p}/X_{\rm H}$. 
BH binaries are allowed to coalesce once their separation is less than $4\Delta x$, where $\Delta x$ is the cell size, and their relative velocity is smaller than the escape velocity of the binary.
BHs grow by smoothly accreting gas from their surroundings according to the boosted \citep{Booth09} Bondi-Hoyle-Lyttleton accretion rate, $\dot M_{\rm BH}=(1-\epsilon_{\rm r})\dot M_{\rm BHL}$, where $\dot M_{\rm BHL}=\alpha 4\pi\bar \rho G^2M_{\rm BH}^2/(\bar u+ \bar c_{\rm s})^{3/2}$. The maximum allowed accretion rate is capped at the Eddington limit, $\dot M_{\rm Edd}=4\pi G M_{\rm BH}m_{\rm p}/(\epsilon_{\rm r}\sigma_{\rm T} c)$, where $c$ is the speed of light, $\sigma_{\rm T}$ is the Thomson cross-section, and $\epsilon_{\rm r}$ is the spin-dependent radiative efficiency. The quantities marked with bars are kernel-weighted as a function of the local gas properties \citep[see][for more details]{Dubois12}.
The so-called AGN feedback from BHs is delivered through a dual jet-heating mode at low, $\chi\le0.01$, and high, $\chi>0.01$, Eddington ratio, $\chi=\dot M_{\rm BH}/\dot M_{\rm Edd}$ \citep{Dubois12}. This is done to mimic the expected behavior of radio jets and quasar winds respectively. 

The coupling efficiency of thermal feedback is set to be $\epsilon_{\rm f,h}=0.15$, in order to reproduce the $M_{\rm BH}$-$M_\star$ relation \citep{Dubois12, Volonteri16}.
For the total energy released in the thermal mode both the coupling efficiency and the spin-dependant radiative efficiency intervene~\citep{Dubois14spin}, that is $\dot E_{\rm BH,h}=\epsilon_{\rm r}\epsilon_{\rm f,h}\dot M_{\rm BHL} c^2$.
The spins of BHs are self-consistently evolved by binary BH coalescence \citep[following][]{Rezzolla08} and smooth gas accretion \citep[see][for details]{Dubois14spin}. 
A geometrically thin and radiatively efficient \cite{Shakura73} disc model is assumed for the quasar mode.  
For the jet mode, we follow the magnetically choked accretion flow solution for the BH spin evolution (always spinning down) of~\cite{McKinney12}.
The energy coupling efficiency of the jet AGN mode ($\dot E_{\rm BH,j}=\epsilon_{\rm f,j}\dot M_{\rm BHL} c^2$) is also given by~\cite{McKinney12}, and follows a U shape with a minimum efficiency at a few per cent for non-rotating BHs, and maximum efficiencies close to 100\% for maximally spinning BHs (see \citealp{dubois20} for details).
Jets in the radio mode deposit momentum and energy into a bipolar outflow (with no opening angle), and redistribute mass with a constant mass loading factor of $\eta=\dot M_{\rm J}/\dot M_{\rm BH}=100$.

Therefore, this model of AGN feedback is similar to that from \texttt{NewHorizon} \citep{volonterietal20,dubois20}, though employed at a different spatial resolution, but differs from \texttt{Horizon-AGN}~\citep{Dubois2014,Dubois2016} because of the modelling of BH spins.

\subsection{Star formation}\label{sub:formation}

A resolved model of star formation would require far higher resolution than is currently possible for such a large volume simulation. We therefore use the statistical approach described in \citet{Rasera06}. Star particles are spawned in gas cells with number density $n_\mathrm{g}>n_\mathrm{0}$. The number density threshold is initially comoving, such that stars may only form in gas cells with an overdensity exceeding $55 \rho_\mathrm{critical}$, where $\rho_\mathrm{critical}$ is the cosmological critical density. This criterion later transitions (at $z\approx21$) to a simple physical density threshold of $n_0=0.1\,{\rm H\,cm^{-3}}$. To prevent the unphysical formation of stars from hot gas, grid cells with a temperature higher than 2,000$\,{\rm K}$ are ineligible to spawn stars. Stars are also forbidden from forming outside the zoom region shown in Figure~\ref{fig:zoom}.

Particles are formed with a mass equal to,
$$m_* = 0.2N_*\frac{\Omega_\mathrm{b}}{\Omega_{m}}0.5^{3l_\mathrm{max}}\,,$$
where $l_\mathrm{max}$ is the maximum refinement level. $m_*$ is in a code unit in which the total sum of matter in the entire volume is set to unity. $N_*$ is an integer factor determined stochastically from a Poisson distribution, such that the average star formation rate obeys 
$$\dot{\rho_*} = \epsilon_* \rho_\mathrm{g} \sqrt{\frac{32 G \rho_\mathrm{g}}{3 \pi}}\,,$$
where $\dot{\rho_*}$ is the star formation rate, $\rho_\mathrm{g}$ the gas density and $\epsilon_* (=0.02)$ is the star formation efficiency parameter. This stochastic calculation of particle masses can lead to inadvertent excessive gas depletion in some grid cells. To prevent this, no more than 90\% of the gas in a grid cell may become a star particle. 

To reduce computational load, the detailed properties of each star particle are written to a file at their formation, and only those properties required for tracking their location or calculating future feedback episodes (position, velocity, initial and current mass, formation time, and metallicity) are preserved in memory. An index number allows a given star particle to be matched to these birth properties.

\subsection{Chemistry}\label{sub:chemistry}

\begin{figure}
    \centering
    \includegraphics[width=1\columnwidth]{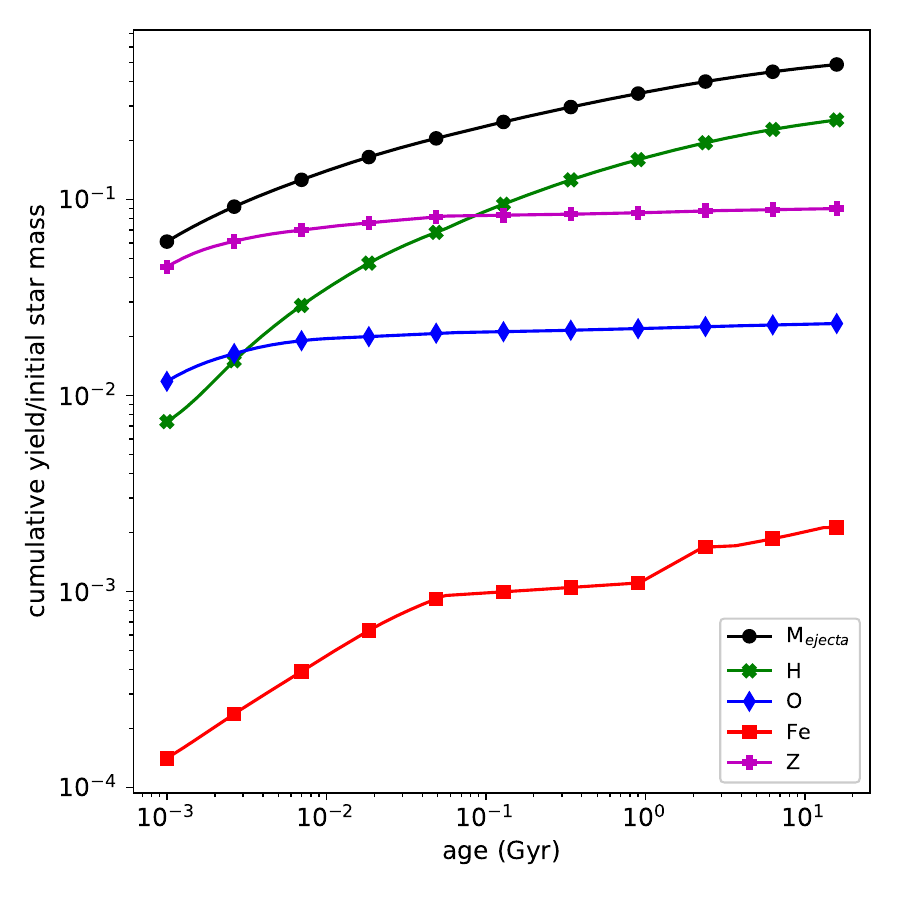}
    \caption{Cumulative mass yield produced by a star particle per unit mass of stars formed as a function of age. The total ejecta mass due to SNII, SNIa and AGB winds is show as black circles, hydrogen mass as green crosses, oxygen in blue triangles, iron as red squares and total metal mass as purple pluses. The yields shown are for a star particle with solar metallicity, other metallicities vary.
    \label{Fig:yieldvsage}
    }
\end{figure}

Stars form in the gaseous phase of the ISM, which interacts with gas reservoirs external to the host galaxy. Depending on their mass, stars may end their lives exploding in an energetic event carrying a particular chemical signature. The elements they release enrich the surrounding gas, and that gas is then used to form the subsequent generations of stars. The observed abundances of stars are thus defined by the properties of the gas from which they were formed, and so preserve a useful record of the history of a galaxy. Therefore, while the simulation retains a record of the physical assembly history of galaxies, and the ensuing star formation, we can gain further insight into the chemical signature of these events by examining the relative abundances of two elements formed by different sources. By following oxygen, which is predominantly formed by SNII, and iron, which is also produced by SNIa, we have access to a `cosmic clock' which can be compared with observations. 

Chemical evolution is modelled following the method described in \citet{Few12}, wherein the production and pollution of the gas phase by stars is based upon a pre-generated yield table. This table dictates the number of each supernovae (SN) type, and the quantity of elements released by SN and Asymptotic Giant Branch (AGB) stars as a function of age and initial metallicity for each stellar population. The pre-generated yield table spans a range of initial stellar metallicities from $3.6\times10^{-5}$ to $50\,{\rm Z_\odot}$, and ages up to $15\,{\rm Gyr}$. It tabulates the number of type-Ia and type-II SN, the total mass of gas as well as the respective masses of H, O, Fe and total of all metals produced for a given metallicity and age. 

Each star particle is treated as a simple stellar population containing individual unresolved stars following an initial mass distribution after \citet{Chabrier03}. Massive stars ($8-100\,{\rm M_\odot}$) are assumed to evolve into SNII with elemental yields for such stars taken from \citet{Kobayashi06}.
Stars with masses of $0.1-8\,{\rm M_\odot}$ evolve into the AGB phase, and deposit elements into the gas phase with elemental yields taken from \citet{Karakas10}. The elemental yield models used are grids, with discrete masses and initial metallicities. For a star with arbitrary mass and metallicity we linearly interpolate yields between neighbouring grid points. To extend the mass and metallicity range to cover all cases that may arise in the main simulation we also extrapolate the tables. For metallicity we simply take the values of the nearest grid point in metallicity, while for the mass we scale the yields of the nearest mass point to the new stellar mass. All interpolations and extrapolations are checked to ensure that total mass is conserved.

Elemental yields for SNIa are taken from \citet{Iwamoto99}. SNIa take place on a considerably longer time scale, governed by an extended time delay distribution over which SNIa release metals back into the ISM. In contrast to SNII, which release metals back into the ISM on a time scale of $10-100\,{\rm Myr}$, SNIa release their metals on a Gyr scale. Our SNIa model is motivated by \citet{Hachisu99}, as employed in \citet{Few14}. SNIa progenitors are treated as binary stars, with primaries in the mass range from $m_{\rm P,l}=3\,{\rm M_\odot}$ to $m_{\rm P,u}=8\,{\rm M_\odot}$ and secondaries that are either main sequence stars $m_{\rm MS,l}=1.8\,{\rm M_\odot}$ to $m_{\rm MS,u}=2.6\,{\rm M_\odot}$ or red giants, $m_{\rm RG,l}=0.9\,{\rm M_\odot}$ and $m_{\rm RG,u}=1.5\,{\rm M_\odot}$. The fraction of secondary companion stars that gives rise to a SNIa progenitor is $b_{\rm MS}=0.05$ and $b_{\rm RG}=0.02$ for main sequence and red giant companions respectively \citep{KAWATA03}. The number of SNIa that have exploded at a time corresponding to a main sequence turn-off mass of $m_\mathrm{TO}$ is given by

\begin{align*}
\label{eq:IaK}
  N_\mathrm{SNIa}(m_\mathrm{TO}) &=          M_0\int_{m_{\mathrm{P,u}}}^{m_{\mathrm{P,l}}} \phi(m) {\mathrm{d}}m \nonumber\\
&\times \Biggr[b_\mathrm{MS} \frac{\int_{{\mathrm{MAX}}(m_{\mathrm{MS,l}},m_{\mathrm{TO}})}^{m_{\mathrm{MS,u}}}\phi(m) {\mathrm{d}}m}{\int_{m_{\mathrm{MS,l}}}^{m_{\mathrm{MS,u}}} \phi(m) {\mathrm{d}}m} \\
&+ b_\mathrm{RG}\frac{\int_{{\mathrm{MAX}}(m_{\mathrm{RG,l}},m_{\mathrm{TO}})}^{m_{\mathrm{RG,u}}}
\phi(m){\mathrm{d}}m}{\int_{m_{\mathrm{RG,l}}}^{m_{\mathrm{RG,u}}}
 \phi(m){\mathrm{d}}m}   \Biggr]\,, \nonumber
\end{align*}
where $\phi(m)$ is the initial mass function of the stellar population with initial mass $M_0$.

The pre-generated yield table stores the cumulative quantities of these values as a function of stellar population age, shown graphically in Figure~\ref{Fig:yieldvsage} for a star particle of solar metallicity. During the simulation, the current age and metallicity of each star particle are used to linearly interpolate between entries in the yield table to calculate the total number of SN that have exploded in that particle and the mass of each element that has been produced so far. The same procedure is also followed using the age of the particle at the time when it last produced feedback. The difference in the values at these two ages constitutes the number of SN or total mass of an element to be released into the ISM at the current time step. In this way, a particle can potentially produce feedback in every time step. To reduce the computational load, and prevent a situation where less than a single supernova explodes, a tolerance is applied, and a star particle must wait until it has accumulated enough SN energy, or mass, to be allowed to contribute to the current time step feedback budget. In test runs this tolerance parameter had little impact on the properties of galaxies, but reduces the CPU time spent in feedback routines.

\subsection{SN feedback}\label{sub:feedback}

Stellar feedback takes the form of passively deposited AGB winds and far more energetic supernovae (types-Ia and -II). Each SN creates an amount of energy which is coupled to the gas phase, either in kinetic or thermal modes. The amount of energy per supernova, chosen to mimic the aggregated energy of core-collapse, superluminous SNIa and pair instability type-II supernovae, is $2\times10^{51}\,{\rm erg}$. During the simulation each particle is compared to the yield table to determine what feedback mode is required, how many SN explode, and the mass of matter to be returned to the gas phase. Parameters controlling the feedback type, and strength, were the subject of an extensive calibration detailed in Appendix~\ref{sec:simparamtuning}. 

In all the \texttt{HR5} runs, if the particle is sufficiently young then it generates SNII in a kinetic mode, depositing $f_\mathrm{ek}(=0.3)$ of the total energy from SNII as kinetic energy, and the remainder as internal energy. To include the effect of SN ejecta sweeping up mass we apply a wind-loading factor to add mass to the resulting blast wave after the method described in \cite{Dubois2008}. The mass of material swept up by the shock wave is a factor of $f_w(=3)$ times the ejected mass from the collective SNII in the star particle. A SN blast wave is imposed across a spherical volume with a radius of twice the one dimensional size of the most refined gas cells.

Once a star has aged sufficiently that all SNII progenitors have been exhausted we switch to generating SNIa and AGB winds. The number of particles eligible to generate this kind of feedback quickly increases to a degree that is impractical to apply to the more computationally costly kinetic feedback method, and so the feedback energy is simply deposited as thermal energy into the nearest grid cell, along with the amount of mass lost. 


\begin{table*}
    \centering
    \caption{The list of variables in snapshots. The particle outputs include dark matter, stars, and cloud particles (which indicate BH particles and the precursors of the BH particles). Additional star particle properties are stored separately as described in the \emph{New Star} column. Variables are dumped in code units so unit conversion is needed using the conversion factors given at each snapshot.}
  \begin{tabular}{m{0.02\textwidth}>{\centering}m{0.15\textwidth}>{\centering}m{0.15\textwidth}>{\centering}m{0.15\textwidth}>{\centering}m{0.15\textwidth}>{\centering\arraybackslash}m{0.15\textwidth}}
    \hline
    \hline
     & Hydro & Particle & New Star & BH & Gravity \\
      \hline
      1 &Position (3D) \footnote{Derived from the central position of a parental grid} & Position (3D) & Position (3D)\footnote{Position of the parental cell} & Position (3D) & Cell potential \\     
      2 &Velocity (3D) & Velocity (3D) & Velocity (3D)\footnote{Velocity of a parental cell} & Velocity (3D) & Gravitational force (3D) \\
      3 &   Cell size \footnote{Derived by $1/2^l$, where $l$ is the grid level}   & Mass   &  Mass & Mass & \\
      4 & Density & ID\footnote{Cloud particles have negative ID numbers, dark matter and star particles has positive ID numbers}    &  ID   & ID  & \\
      5 & Thermal pressure   & Grid Level  & Grid Level  &  Formation epoch & \\
      6 & Metallicity (Z)  & Potential    & Birth epoch & Bondi accretion rate  &\\
      7 &$f_{\rm H}$ (X) & Birth epoch\footnote{Dark matter and cloud particles have birth epochs equal to zero}  & Parent cell density & Eddington accretion rate & \\
      8 &$f_{\rm O}$ & Metallicity (Z)  & Parent cell temperature & Gas spin axis &\\
      9 &$f_{\rm Fe}$ & Initial mass & Metallicity (Z) & BH spin axis  &\\
      10 &ID on cpu \footnote{Given as the location on a particular cpu, therefore non-unique across the whole snapshot, and not consistent over time}   &              & $f_{\rm H}$ (X) & BH spin amplitude & \\
        11 &           &              & $f_{\rm O}$ & BH efficiency &\\
        12 &          &              & $f_{\rm Fe}$ & BH radiative efficiency &\\
        13 &          &              &              & Amount of feedback energy &\\
    \hline
    \end{tabular}
  \label{tab:variables}

\end{table*}

\section{Output Data}\label{sec:output}
In this section, we briefly describe the output data produced by the \texttt{HR5} runs: first the snapshots (Sect.~\ref{sub:snap}), then the lightcone (Sect.~\ref{sub:lightcone}), and finally the clusters (Sect.~\ref{sub:clusdata}).

\subsection{Snapshot Data}
\label{sub:snap}
The snapshots of \texttt{HR5} contain the information of all particles (dark matter, stars, BHs), the AMR grid structure, as well as the gas properties and gravitational potential in each grid cell. The properties of the new stars that have formed since the latest snapshot was produced are also recorded. The output variables are listed in Table~\ref{tab:variables}.

In the initial simulation design, the total number of snapshots was set to 171, from $z=200$ to $z=0$. The first 21 snapshots are uniformly spaced on a logarithmic scale of the expansion factor, from $z=200$ to 10. The remaining 150 snapshots are set in the same manner from $z=10$ down to $z=0$. The redshifts of the snapshots can be slightly different from this initial setup, because {\tt RAMSES} produces snapshot data at the beginning of the subsequent main time step (coarse step). This is because the different refinement levels have different time step sizes and it is only at the coarse step that all the levels are synchronized.  In practice, the latest snapshot is at a redshift of $z=0.625$, and the total number of snapshots is 147. The first snapshot ($z\sim200$) has a size of $\sim$2~TB, but as the system evolves over time, the snapshots at $z<1.5$ eventually reach a size of $\sim$10~TB, and include $\sim 10^{10}$~particles and more than $4\times10^{10}$~cells.

Particle and cell data is stored in chunks which correspond to spatially contiguous regions defined by the {\tt RAMSES} peano-hilbert curve used to decompose the simulation volume. This differs from the \texttt{Illustris} simulation \citep{Nelson2015} which order their snapshot data based on halo mass. This is of course dependent on the particular problems the simulations are expected to address. Much of this information, is however, also stored in the halo catalogue discussed in Sect.~\ref{sec:finding}.

\subsection{Lightcone Data}
\label{sub:lightcone}

\begin{figure*}
    \centering
    \includegraphics[width=\textwidth]{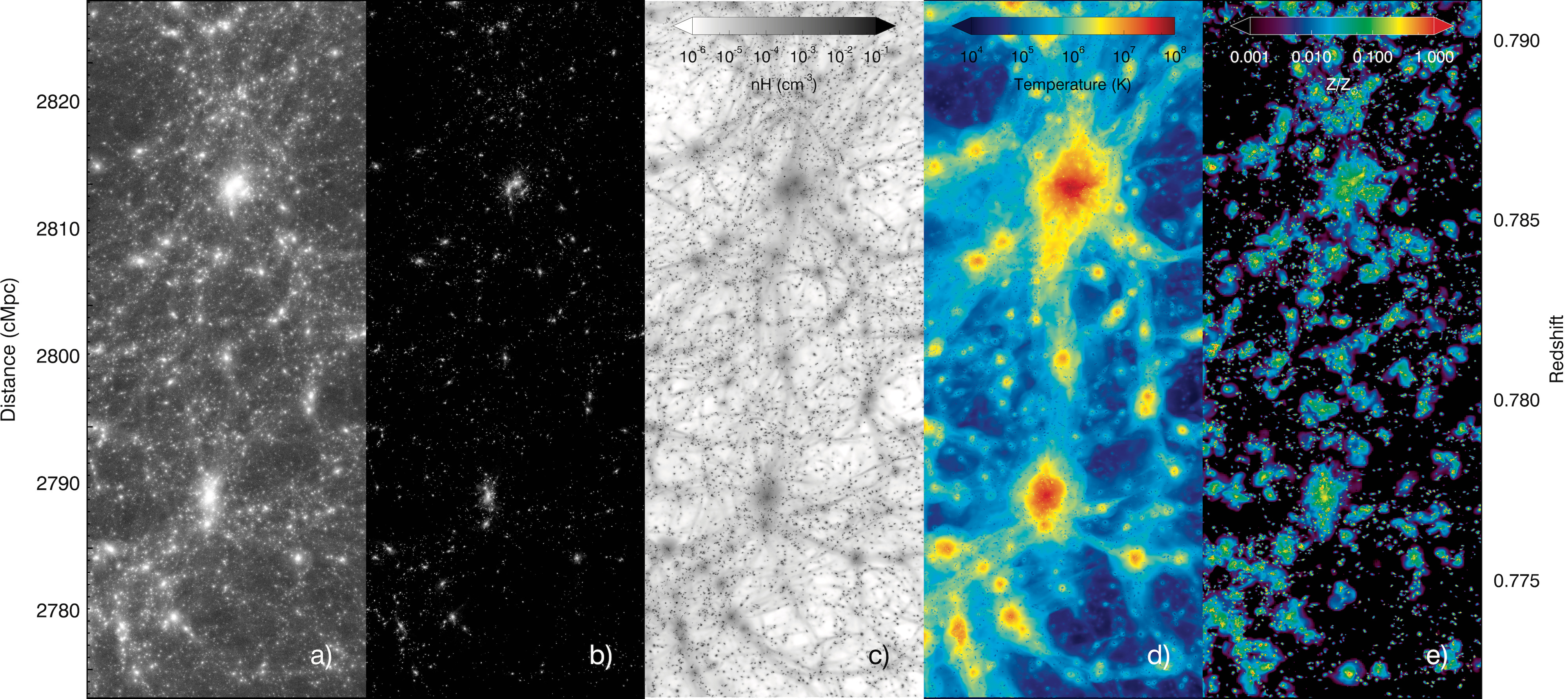}
    \caption{Projections of a) dark matter mass, b) stellar mass, c) gas density, d) gas temperature, and e) gas metallicity in a region of the lightcone at $z\sim0.78$. The observer plane is located past the bottom of the images.
}
    \label{fig:lightspace}
\end{figure*}

In order to compare with redshift surveys where redshift changes continuously with distance, we need to construct lightcone data. The geometry of the high resolution region of the \texttt{HR5} simulations requires us to generate lightcones with a small opening angle, thus generating mock pencil beam surveys.
We generated a pair of past lightcone space data on-the-fly using the far-field approximation, starting from virtual observers located on the surface of the periodic boundary and looking in opposite directions along the long axis of the zoom region. A key feature of this choice is that the virtual observers at $z=0$ measure distances to cells and particles on the surface parallel to the observing plane at any given redshift. The geometry of this data is thus a long cuboid, instead of the traditional spherical sector of observed lightcones. Therefore, celestial objects located at the same distance from an observer at a given location inevitably have slightly different redshifts. This is however almost negligible at high redshift. Indeed, \citet{nishioka99} show that the far-field approximation does not notably affect the two-point correlation statistics at high redshifts. An advantage of this approach is that we can minimize missing cells between lightcone slices at $t$ and $t+\Delta t$, which is otherwise unavoidable in generating lightcone space data from simulations based on cubic mesh cells \citep[see][]{gouin19}. All the variables in this data are the same as those in the snapshots, but the full AMR structure is not stored and the gas variables are recorded only for leaf cells. Figure~\ref{fig:lightspace} shows examples of projections of a lightcone region for dark matter, stars, gas density, gas temperature, and gas metallicity (from left to right).

\subsection{High output cadence in the densest regions}\label{sub:clusdata}
At every coarse time-step of our simulations we stored data for five spherical volumes in the zoomed region that are expected to form massive clusters at $z=0$. Down to $z=0.625$, a total of 758 snapshots were stored for each region, which provides time resolution roughly five times better than that of the snapshots. To find the over-density regions that end up forming clusters at $z=0$, we first carried out a low resolution run in which the zoomed region is refined from level 11 up to level 15 (corresponding to a maximal resolution of $\Delta x\sim32\,\kpc$). We then identified halos in the zoomed region at $z=0$, and picked the five most massive ones. Because the zoomed region of the low resolution run encloses a volume larger than that of our main run, we only chose halos that are fully located inside the zoom region at a refinement level above 13. To minimize boundary effects and contamination by high mass particles flowing from low resolution regions, we ensured that all the spherical regions are at least $20\,\cMpc$ away from the boundary of the zoomed region. We then back-traced all the collision-less particles forming the halos in the initial conditions ($z\sim200$). We identified the sphere of the smallest radius which enclosed all the particles at this epoch. All particle and gas properties in leaf cells were then recorded from each of the five spherical regions at every main time step during all the \texttt{HR5} runs. This allows us to investigate the formation and evolution of massive galaxy clusters with a time resolution much finer than that of the snapshots. The masses of the target halos identified in the low resolution simulation are listed in the first row of Table~\ref{tab:cluster}, along with the ones measured in the latest snapshot ($z=0.625$) of the main run (\texttt{HR5}).

\section{Halo \& Galaxy Identification}  
\label{sec:finding}

In this section we present the methods implemented to identify virialized halos and galaxies in the \texttt{HR5} run, first using a FoF algorithm (Sect.~\ref{sub:fof}), and then present an improved approach (Sect.~\ref{sub:PSBGF}), before describing the final  galaxy catalogue (Sect.~\ref{sub:galaxycat}).  Finally, Sect.~\ref{sub:clusters} focuses on the properties of clusters. 

\subsection{FoF Halo Identification}
\label{sub:fof}
{\tt RAMSES} produces data for three types of particles representing dark matter, stars, and BHs, along with data for the gas cells. The gas cells, unlike the particles, record the mean density, from which we can calculate the mass contained in the cells, given their refinement level. To identify virialized regions, we transform the gas mass in the cells into ``pseudo'' gas particles, and apply a percolation method to identify virialized structures.

We used an extended Friend-of-Friend (FoF) method to identify virialized halos with a variable linking  length of
\begin{equation}
    	\!\!l_{\textrm{link}} =\!\! 0.2 \times\! \left({ m_p \over \Omega_{m0}\varrho_{c} }\right)^{1/3},
\end{equation}
where $\varrho_{c}$ is the critical density at $z=0$, and $m_p$ is the particle mass. This variable FoF scheme is applied to the mixture of particles of dark matter, star, BH, and gas. 
To link two particles of different mass,
we took the average linking length: $l_\textrm{comb} =\left( l_\textrm{1} + l_\textrm{2}\right)/2$.
With this chain of linkages we can detect a group of multi-species particles.


\begin{figure*}
    \centering
    \includegraphics[width=0.9\textwidth]{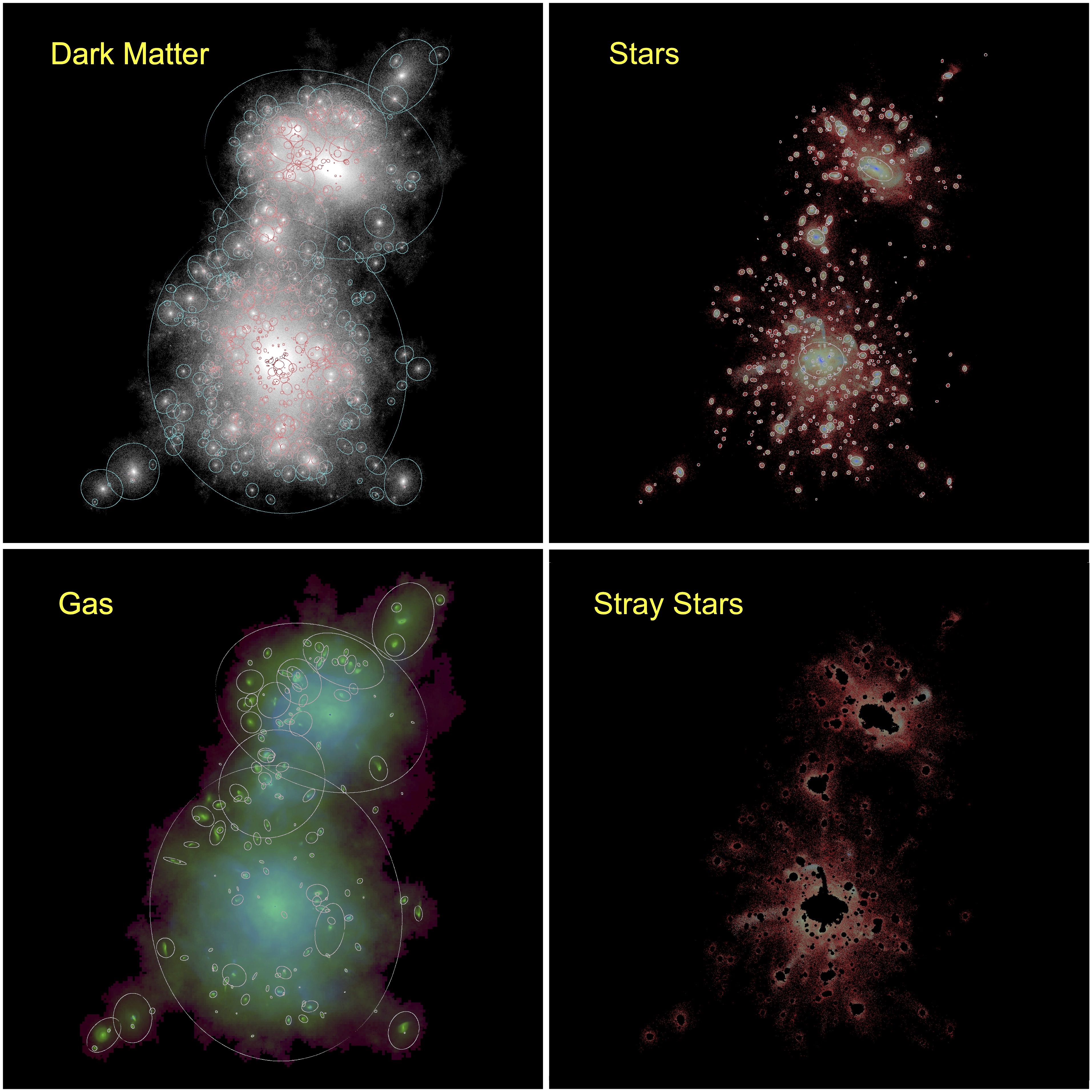}
    \caption{An example of FoF halo and galaxy findings applied to a halo having a total mass of $M_\textrm{tot}=4.97\times10^{14}\Msun$ with the individual mass components of $M_\textrm{DM}=4.76\times10^{14}\Msun$, $M_\textrm{*}=6.56\times10^{12}\Msun$, and $M_\textrm{gas}= 4.62\times10^{13}\Msun$, respectively. Counter-clockwise from the top-right panel shown are the distributions of stars, dark matter, and gas cells with projected ellipsoids which are fitted to each component of galaxies. Also the distribution of cluster stray stars is shown in the bottom-right panel being masked out by the stellar components of galaxies.
    }
    \label{fig:psb_gal}
\end{figure*}

\begin{table}
    \centering
    \caption{Several important figures of \texttt{HR5} at $z=0.625$. The number of particles, leaf cells, FoF halos, and galaxies are listed. $M_{\rm tot}$ is the total mass of all particles and cells in an individual FoF group.}
    \begin{tabular}{l|r}
    \hline
      Object   & Number \\
       \hline
     Dark matter particles & 7,774,614,016  \\
     Star particles  & 2,210,233,512 \\
     BH particles & 899,562 \\
     Leaf cells & 42,342,076,008 \\
     FoF halos ($M_{\textrm{tot}}>10^{11}\,\Msun$) & 184,956 \\
     Cluster halos ($M_\textrm{tot} >10^{14}\, \Msun$) & 102 \\

     Galaxies ($M_*>10^9\,{\rm M_\odot}$)  & 290,086 \\
    \hline
    \end{tabular}
    \label{tab:number}
\end{table}


\subsection{PSB-based Galaxy Finder (\texttt{PGalF})}\label{sub:PSBGF}

A new galaxy finding algorithm was developed to search substructures from FoF haloes for satellite galaxies,
which are gravitationally self-bound and tidally stable in group or cluster environments. This finder is based on the original subhalo finder, \texttt{PSB}, which was developed to identify subhalos inside dark-matter-only FoF halos \citep{Kim06}. This approach of identifying FoF halos and subsequently the substructures within them is conceptually similar to the well known \texttt{subfind} algorithm \citep{Springel2001}. 

Here we are able to identify {\it galaxies}, which have internal and external properties different from dark matter subhalos in several respects. Stellar components tend to be more compact than their dark matter counterparts and thus better survive the strong background tidal force exerted by their host halo. The background potential, however, is largely governed by the dark-matter component. Sometimes, satellite galaxies may even lose their associated dark matter component through tidal disruption as they orbit within the group.  In our search for galaxies we thus start from the stellar distribution, which is then used as a seed to explore the distribution of the other particle species.

\begin{figure*}
    \centering
    \includegraphics[width=\textwidth]{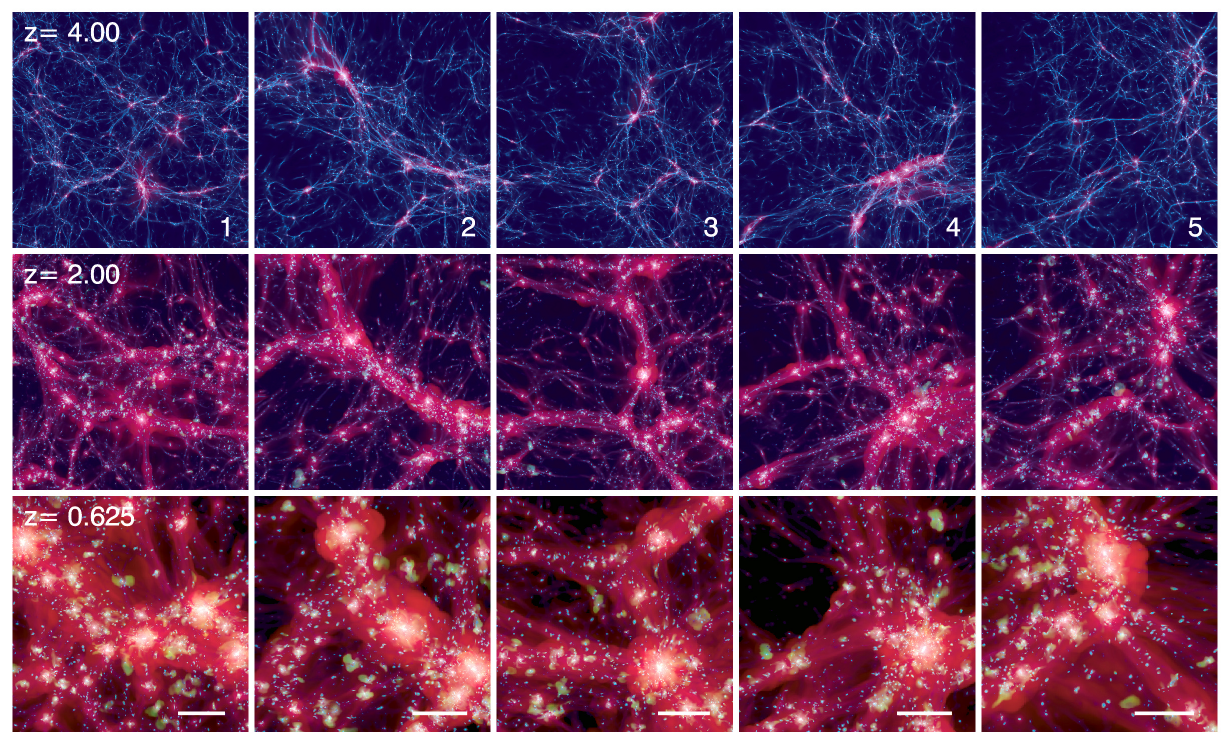}
    \caption{Composite images of the five regions expected to end up enclosing the most massive clusters located in the zoomed region at $z=0$. Blue, red, green, and white colors represent gas density, gas temperature, metallicity, and stellar mass density, respectively. The five columns are for the five different regions at $z=$4.0, 2.0, and 0.625 (first, second and third row respectively, as indicated in the first column panels). The white horizontal lines in the bottom panels denote a scale length of $5\,\cMpc$.}
    \label{fig:clusters}
\end{figure*}

We first identify the $N_{\rm s}$ nearest star particle neighbors of each star particle and measure local densities using the $W_4$ smoothed particle hydrodynamics (SPH) density kernel \citep{Monaghan85}. 
We build a neighborhood network using chains of $N_n$-nearest neighbors at each particle position. This means that all particles have their own neighboring connections. This coordinate-free method suppresses the ambiguities arising in building the density field, and reduces the number of parameters required by the galaxy finding algorithm.

Secondly, we search for the peaks in the stellar mass density field at each particle position. These peaks are local maxima with respect to their own $N_n$ neighbors.
To alleviate the identification of spurious local peaks due to Poisson noise  we adopt two approaches. One is to merge multiple peaks if their distance is less than $l_\textrm{merge}$. The other is to set both the minimum number of stars, and minimum stellar mass, in the core region. Here a core region is a volume identified by progressively lowering the density threshold down to the point where
the isodensity surface encloses another peak. This is similar to the watershed method. At this point, particles within the isodensity surface are considered to belong to a galaxy candidate centered on the enclosed peak.

Third, after extracting core particles related to a density peak, we apply density cuts to group the remaining (non-core) particles utilizing the watershed method. Around each density core we search for non-core particles whose densities are between the $i\,^{\rm th}$ and $i+1^{\rm th}$ density thresholds. We gather those particles to build a ``shell'' group. By definition, a shell group should surround other density groups.
Accordingly all particles are split into core and shell groups. 

Finally, to complete the membership decision we check the tidal boundary of each galaxy candidate, and total energy of particles within the boundary. Even though a particle may be bound to a galaxy, it is not considered a member unless it is within the tidal radius.

Figure~\ref{fig:psb_gal} shows the most massive halo in \texttt{HR5} identified using the variable FoF approach at $z=0.76$. Even with varying particle mass, the derived distributions of dark matter (top-left) and gas (bottom-left) are almost the same. The panels for DM and gas show spurious bridges between two most massive substructures, which clearly demonstrates one of the well known problems of the FoF method in detecting overdense structures \citep[e.g.][]{Klypin11}. This means that the halos identified using FoF contain not only virialized structures, but also collections of virialized objects linked by filaments.
We also delineate the galaxies identified by the \texttt{PGalF} (\texttt{PSB}-based Galaxy Finder) with projected ellipsoidal shapes in Figure~\ref{fig:psb_gal}. 

While the dark matter and gas cells are widely distributed and match well with each other, the stellar components form structures much clumpier than the dark matter and gas distribution. In the bottom-right panel, we also add the image of the distribution of stray star particles which are not bound to any galaxies in the cluster. We mask out the projected regions of galaxy stellar components (top-right panel) to better present the diffuse stellar features (merging streams) in this cluster. The color gradient changing from red, green, to blue marks increasing stellar metallicity. The color gradient thus represents the metallicity gradient in intra-cluster light \citep [e.g.][]{montes18}. We plan to give a comprehensive description of the \texttt{PGalF} in a series of subsequent papers, along with an in-depth comparison to other galaxy finders.

\begin{table*}
    \centering
    \caption{The properties of the most massive halos and galaxies in the 5 cluster regions at $z=0.625$ stored with time resolution finer than that of the main snapshots. $M_{\rm tot}$ is the total mass of all particles and cells contained within the FoF halos. $M_{\rm enclosed}$ is the mass of all the matter contained within the expected radius of the object at $z=0$. $M_{200c}$ is measured from the density peak of each FoF halo to the radius at which mean matter density is 200$\rho_{\rm crit}$. The BCG is defined to be the most massive galaxy in the FoF halo. $M_{\rm BH}$, $M_{\rm cold}/M_*$, sSFR, and $M_*$-weighted age show the mass of the SMBH, the mass fraction of cold gas ($T<10^4\,{\rm K}$), the specific star formation rate (sSFR) averaged over last $100\,{\rm Myr}$, and the stellar mass-weighted age of the BCG, respectively. Galaxy stellar mass $M_*$ and cold gas mass $M_{\rm cold}$ are measured inside five times the half mass radius of the total stellar mass bound to galaxies. This spherical aperture is adopted to minimize the loss of bound mass, and reduce any potential contamination by the outskirts of neighbouring galaxies. This cut ensures that more than $99.5\%$ of the bound stellar and cold gas mass is accounted for, on average.}
    \begin{tabular}{cccccc}
    \hline
        & \texttt{Cluster 1} & \texttt{Cluster 2} & \texttt{Cluster 3} & \texttt{Cluster 4} & \texttt{Cluster 5} \\ 
       \hline
     
     $M_{\rm tot}$ at $z=0$\footnote{Note that this mass is the total FoF group halo mass estimated from a low resolution simulation used to identify cluster candidates at $z=0$. (see Sect. \ref {sub:clusters})} (M$_{\odot}$)& $8.2\times10^{14}$ & $7.0\times10^{14}$ & $3.9\times10^{14}$ & $3.9\times10^{14}$ & $3.7\times10^{14}$  \\
      $R_{\rm sphere}$ (cMpc) & 27.5 & 23.5 & 24.7  & 23.4 & 21.7  \\
 \hline  
    
     $N_{\rm halo}\,(M_{\rm tot}>10^{13}\,{\rm M_{\odot}})$ & 41 & 30 & 28 & 16 & 16\\
     $M_{\rm enclosed}$ (${\rm M_{\odot}}$) & $6.4\times10^{15}$ & $4.4\times10^{15}$ & $4.1\times10^{15}$ & $3.2\times10^{15}$ & $2.5\times10^{15}$  \\
        \hline
     Most massive $M_{\rm tot}$ ($M_\odot$) & $2.7\times10^{14}$ & $ 2.4\times 10^{14}$ & $3.3\times 10^{14}$  & $3.4\times10^{14}$ &$2.5\times10^{14}$  \\
     $M_{200c}$ ($M_\odot$) & $2.4\times10^{14}$ & $ 2.2\times 10^{14}$ & $2.9\times 10^{14}$  & $2.1\times10^{14}$ &$1.7\times10^{14}$ \\
     $M_*$ of BCG (M$_\odot$) & $8.0\times10^{11}$ & $ 8.3\times 10^{11}$ & $9.5\times 10^{11}$  & $1.1\times10^{12}$ &$9.2\times10^{11}$  \\
     $M_{\rm BH}$ ($M_\odot$) & $3.2\times10^{9}$ & $ 3.4\times 10^{9}$ & $1.0\times 10^{9}$  & $7.0\times10^{8}$ &$2.7\times10^{9}$  \\
     $M_{\rm cold}/M_{*}$ & 0.0127& 0.0040 &  0.0022 & 0.0020 & 0.0059 \\
     sSFR (yr$^{-1}$) & $3.2\times10^{-11}$ & $5.9\times10^{-12}$ & $8.1\times10^{-12}$ & $7.1\times10^{-11}$ & $5.7\times10^{-12}$ \\
     Age ($M_{*}$-weighted, Gyr) & 4.25 & 5.04 & 4.77 & 5.12 & 4.63\\
      \hline
        2nd massive $M_{\rm tot}$ (M$_\odot$)& $2.1\times10^{14}$ & $ 1.9\times 10^{14}$ & $7.5\times 10^{13}$  & $8.1\times10^{13}$ &$7.3\times10^{13}$  \\
     $M_{200c}$ (M$_\odot$) & $1.3\times10^{14}$ & $ 1.6\times 10^{14}$ & $6.2\times 10^{13}$  & $3.1\times10^{13}$ &$5.2\times10^{13}$ \\
     $M_*$ of BCG (M$_\odot$)&  $9.0\times10^{11}$ & $ 8.4\times 10^{11}$ & $4.5\times 10^{11}$  & $2.9\times10^{11}$ &$4.3\times10^{11}$  \\
     $M_{\rm BH}$ (M$_\odot$) & $5.5\times10^{9}$ & $ 6.7\times 10^{8}$ & $9.0\times 10^{8}$  & $9.8\times10^{8}$ &$2.3\times10^{9}$  \\
     $M_{\rm cold}/M_{*}$ & 0.0108& 0.0815 &  0.1127 & 0.0357 & 0.0297 \\
     sSFR (yr$^{-1}$) & $9.3\times10^{-11}$ & $1.2\times10^{-10}$ & $2.1\times10^{-10}$ & $9.3\times10^{-11}$ & $4.0\times10^{-11}$ \\
     Age ($M_{*}$-weighted, Gyr) & 4.43 & 4.24 & 4.05 & 3.77 & 4.05\\
      \hline
     3rd massive $M_{\rm tot}$ (M$_\odot$)& $1.4\times10^{14}$ & $ 1.0\times 10^{14}$ & $6.9\times 10^{13}$  & $5.7\times10^{13}$ &$7.9\times10^{13}$  \\
     $M_{200c}$ (M$_\odot$) & $1.1\times10^{14}$ & $ 8.0\times 10^{13}$ & $4.8\times 10^{13}$  & $3.5\times10^{13}$ &$5.2\times10^{13}$ \\
     $M_*$ of BCG (M$_\odot$)&  $5.3\times10^{11}$ & $ 5.4\times 10^{11}$ & $3.8\times 10^{11}$  & $3.3\times10^{11}$ &$2.2\times10^{11}$  \\
     $M_{\rm BH}$ (M$_\odot$) & $2.3\times10^{8}$ & $ 1.9\times 10^{9}$ & $8.0\times 10^{8}$  & $1.9\times10^{9}$ &$5.8\times10^{8}$  \\
    $M_{\rm cold}/M_{*}$ & 0.0495& 0.0155 &  0.1221 & 0.0156 & 0.0917 \\
    sSFR (yr$^{-1}$) & $4.4\times10^{-11}$ & $2.9\times10^{-11}$ & $1.4\times10^{-10}$ & $5.3\times10^{-11}$ & $1.0\times10^{-10}$ \\
    Age ($M_{*}$-weighted, Gyr) & 4.15 & 4.57 & 3.69 & 4.65 & 4.01\\
    \hline
    \end{tabular}
    \label{tab:cluster}
\end{table*}

\subsection{Galaxy Catalogue}
\label{sub:galaxycat}
We generated a galaxy catalogue using \texttt{PGalF}. Only galaxies with $M_{*}>10^9 \,{\rm M_{\odot}}$ are listed in our galaxy catalogue, because of the telltale resolution signature of a decline in the galaxy mass functions below that mass. 
Recall that \texttt{PGalF} identifies self-bound objects from FoF groups composed of all matter components: gas cells, DM, stellar, and BH particles. \texttt{PGalF} is thus designed to match the stellar component of galaxies with their FoF halos, subhalos, and BHs.

\begin{figure*}
    \centering
    \includegraphics[width=\textwidth]{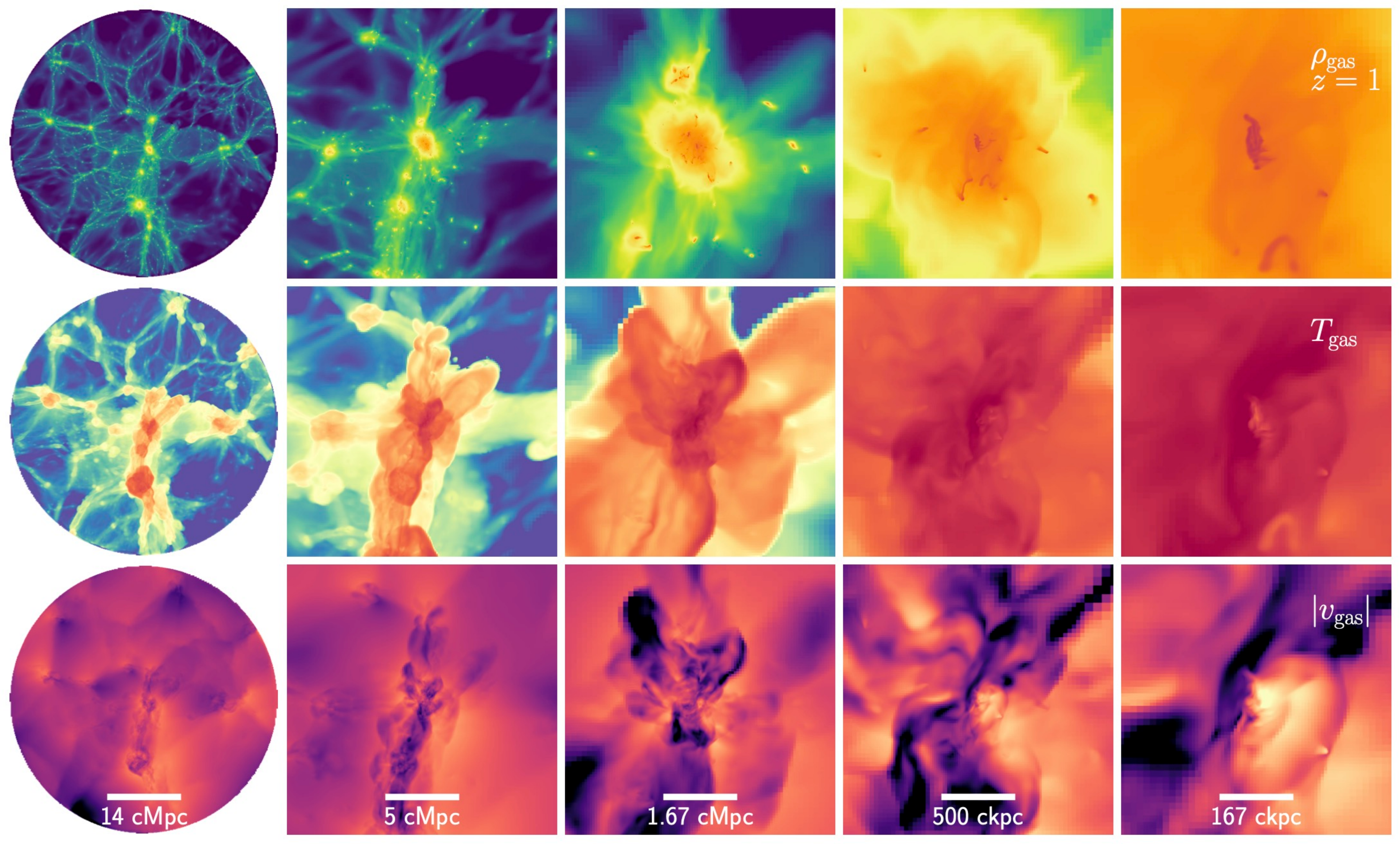}
    \caption{Distribution of the gas density (top row), gas temperature (middle row) and gas velocity (bottom row) in \texttt{Cluster 1} at $z = 1$ on progressively smaller spatial scales. 
    From left to right: width and height are $56~\mathrm{cMpc}$, $20~\mathrm{cMpc}$, $6.67~\mathrm{cMpc}$, $2~\mathrm{cMpc}$, and $0.67~\mathrm{cMpc}$.
    }
    \label{fig:cluster1}
\end{figure*}

Table~\ref{tab:number} lists the matter content in the whole volume. \texttt{HR5} contains  $\sim 7.8\times10^9$ DM particles. By $z=0.625$, $\sim 2.2\times10^9$ stellar particles and $\sim 9.0\times10^5$ of BH particles have been formed, while the gas properties are traced using $\sim4.2\times10^{10}$ leaf cells. From these particles and cells, \texttt{PGalF} identifies $\sim 3.3\times10^6$ FoF groups above $M_{\rm tot}>2\times 10^9\,\Msun$ (corresponding to the mass of 30 DM particles), 
where $M_{\rm tot}$ is the total mass of all particles and cells in an individual FoF group halo. From these, we identified 290,086 galaxies with $M_{*}> 10^9\,\Msun$ at the final snapshot. The center of a galaxy is defined by the density peak of stellar particles only. The kinematics and reduced properties such as angular momentum, half mass radius, rotational velocity, and velocity dispersion of gas and stars in a galaxy are computed from the stellar density peak. We also compute the rest-frame galaxy luminosities in the \texttt{Johnson} and \texttt{SDSS} filter systems based on the ages and metallicities of individual stellar particles. For this, we use the mass-to-light ratios of single-age and single-metallicity stellar populations provided by the \texttt{E-MILES} stellar population synthesis model \citep{vazdekis10,ricciardelli12,vazdekis16}. As mentioned in Sect. \ref{sub:chemistry}, a Chabrier IMF is adopted in \texttt{HR5}. Dust reddening is assumed to be fully corrected in the galaxy catalogue. Directly observable photometric predictions, i.e. redshifted and reddened by dusts, will be provided by Song et al. (in prep).

As seen in Figure~\ref{fig:zoom}, the boundary of the high resolution region becomes bumpy with the evolution of the density field, inevitably being contaminated by low level, high mass particles. For studies which may be affected by potential boundary effects, we also compute distances from galaxies to the surface isolating the whole low level particles in each snapshot. The geometry evolution of the zoomed region is illustrated in Appendix~\ref{sec:geometry}.

The matter content of haloes and their substructures can vary, but down to the latest snapshot, no halo above $M_{\rm tot}=10^9\,\Msun$ and no galaxy above $M_{*}>10^9\,\Msun$ shows a dark matter deficit. This is consistent with \citet{saulder20}, who analyzed \texttt{Horizon-AGN} \citep{Dubois2014}, a cosmological hydrodynamical simulation also using \texttt{RAMSES} \citep{Teyssier02,Dubois12}, and found no such deficient galaxies.

\subsection{Clusters grown in the densest environments}\label{sub:clusters}

We present the evolution of cluster candidates in spherical regions enclosing five of the highest density environments in \texttt{HR5}. As previously mentioned, for these regions we have
saved data much more regularly than the rest of the simulation for which only 147 snapshots are available down to $z=0.625$. Figure~\ref{fig:clusters} shows the composite images of the five cluster candidates at $z=4$, 2, 0.625 in \texttt{HR5}. One can see the growth of structures, galaxies, increasing temperature (reddish) and metallicity (greenish) with cosmic time.

In Table~\ref{tab:cluster}, we show the main properties of our cluster candidates at the latest epoch ($z=0.625$). Massive structures are being assembled in each spherical region with enclosed mass ranging from $2.5-6.4\times10^{15}\,{\rm M_\odot}$, and with between 16 and 41 FoF groups more massive than $10^{13}\,\Msun$. As can be seen in Figure~\ref{fig:clusters}, each cluster exists within a substructure rich environment, and a range of local environment geometries.

The most massive halos have $M_{\rm tot}$ and $M_{200c}\sim 1.7-3.4\times10^{14}\,\Msun$ and galaxies with $M_{*}\sim8\times10^{11}-1.1\times 10^{12}\,\Msun$ at $z=0.625$. The sSFRs show that the most massive galaxies in the most massive halos are almost quenched with $M_{\rm cold}/M_*\lesssim1\%$. To estimate the $z=0$ mass of the FoF halos contained in the spherical region at $z=0.65$, we referred to the \texttt{Horizon Run 4} simulation \citep{2015JKAS...48..213K}. This is a cosmological $N$-body only simulation, covering a cubic volume of $4.37\,{\rm cGpc}$ on a side. We traced the halo merger trees of \texttt{Horizon Run 4} from $z=0.64$ to $z=0$.  We found that halos can easily double their mass between $z\sim0.64$ and $z=0$. For $M_{\rm tot}<10^{14}\,\Msun$ at $z\sim0.64$, only a small fraction of halos show a mass decrease, mainly caused by tidal fields induced by their massive neighbors. This suggests that haloes of $M_{\rm tot}>5\times10^{13} \Msun$ at $z\sim0.63$ have a high chance of forming cluster-scale haloes until $z=0$.

In Figure \ref{fig:clusters}, the top three massive halos in \texttt{Cluster 1} (three clumpy structures on the mid-right side at $z=0.625$) are close to each other ($d < 7\,{\rm cMpc}$) and have relative velocities ($\sim 500\,{\rm km\,s^{-1}}$) enough to merge before $z=0$. The sum of their total FoF group masses is $6.2\times10^{14}\,\Msun$ at $z=0.625$, implying that at least one halo in the zoomed region could potentially build a massive Coma-like cluster with $M_{\rm tot}\sim10^{15}\,\Msun$ by $z=0$. One can see an additional, and even larger, halo ($M_{\rm tot}\sim3.3\times10^{14}\,\Msun$) than the three halos in the outskirts (top left corner at $z=0.625$) of \texttt{Cluster 1}, but it is too distant ($d\sim22\,{\rm cMpc}$) to merge with the three other halos before $z=0$. 

Figure~\ref{fig:cluster1} shows the density (top row), temperature (middle row), and velocity (bottom row) for the gas component in \texttt{Cluster 1} on different scales at $z=1$. Despite its large host halo mass, the central galaxy is fed by filamentary structures of gas, forming colder and denser clouds in their cores. The gas velocity maps in the bottom row reveal the rotational and turbulent motion of such gas clouds around the galaxy. As seen in Figure~\ref{fig:clusters}, galaxies in dense environment gradually see the temperature of their circumgalactic medium rise over time, mainly though AGN feedback which eventually evaporates most of these colder clouds.

\section{Global Properties}  \label{sec:global}
The global statistical properties of \texttt{HR5} are presented in this section. We start by detailing two point statistics for galaxy clustering (Sect.~\ref{sub:structure}). We then present the cosmic star formation history (Sect.~\ref{sub:SF}), stellar mass function (Sect.~\ref{sub:SMF}), gas phase metallicity of galaxies (Sect.~\ref{sub:chem}), the properties of super massive black holes (Sect.~\ref{sub:BH}) and filament and void statistics (Sect.~\ref{sub:cosmicweb}). We note that all the galaxy properties are measured inside five times the half mass radius $R_{1/2}$ of the stellar mass bound to galaxies. This spherical aperture is adopted as a compromise between maximizing the amount of bound mass that is taken into account in our measurements, and reducing potential contamination from the outer regions of neighbouring galaxies.  This aperture contains $\sim99.5\%$ of bound stellar and cold gas mass on average. This mass measurement scheme is fundamentally different from that used in observations. Thus, this should be kept in mind when a comparison is made between the \texttt{HR5} galaxies and those of observations. More details are given in Sect.~\ref{sub:SMF}.

\subsection{Matter Power Spectra \& Clustering at the base level} \label{sub:level8}
\begin{figure}
    \centering
    \includegraphics[width=1\columnwidth]{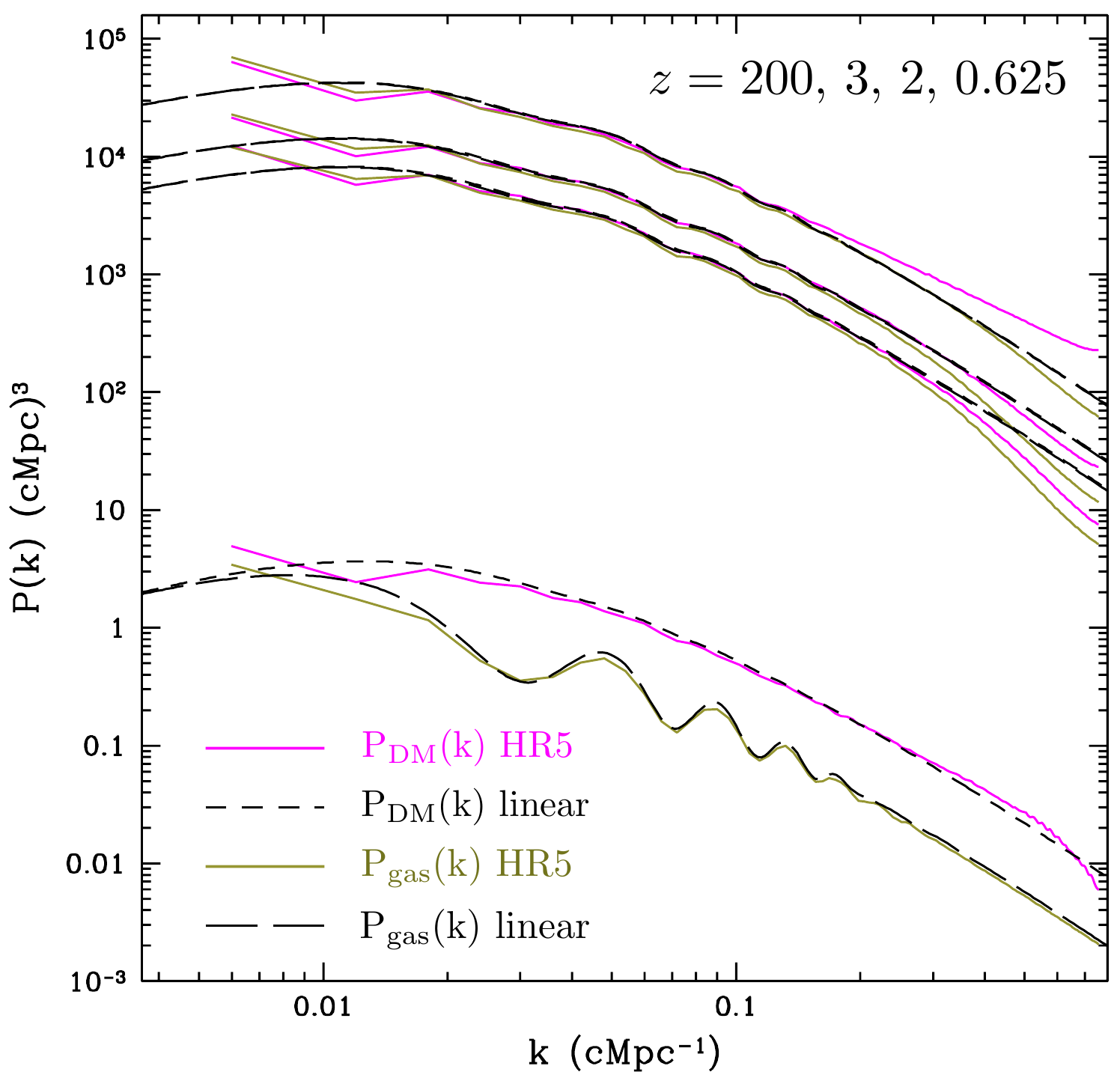}
    \caption{Power spectra measured in $256^3$ root cells of \texttt{HR5} (level 8) and the linear model. The power spectra of dark matter (magenta) and gas (olive) in \texttt{HR5} are shown with their corresponding linear predictions (short-dashed and long-dashed black curves, respectively) at redshifts $z=200$, 3, 2, and 0.625, from the bottom set of the lines to the top. We obtained the linear predictions from the \texttt{CAMB} package \citep{lew00}. Note that the scale of the first dip of the baryonic wiggle is fully contained within the volume of \texttt{HR5}.}
    \label{fig:pkb}
\end{figure}
The power spectrum of the matter density field has widely been used to test the reliability of simulations. Even though the small-scale nonlinear features emerge at low redshifts, the overall shape of the simulated power spectrum is roughly preserved.
In Figure~\ref{fig:pkb}, we show the power spectra of two matter species, i.e. dark matter particles and gas cells, at the coarse level of \texttt{HR5} with $\sim4.09\,\cMpc$ pixels. The power spectra are measured from the mass density field assgined on a uniform grid of $256^3$ voxels. At the starting redshift ($z=200$), the initial condition of the matter distribution accurately follows the linear theory. The baryonic wiggle is also well captured, and the amplitude difference between the power spectra of dark matter and baryons is correctly reproduced in the initial condition of \texttt{HR5}. During the evolution, the amplitude of \texttt{HR5} power spectrum in good agreement with the linear theory, except on small scales ($k \gtrsim 0.1~{\rm cMpc}^{-1}$), where it grows faster for DM.

\begin{figure}
    \centering
    \includegraphics[width=1\columnwidth]{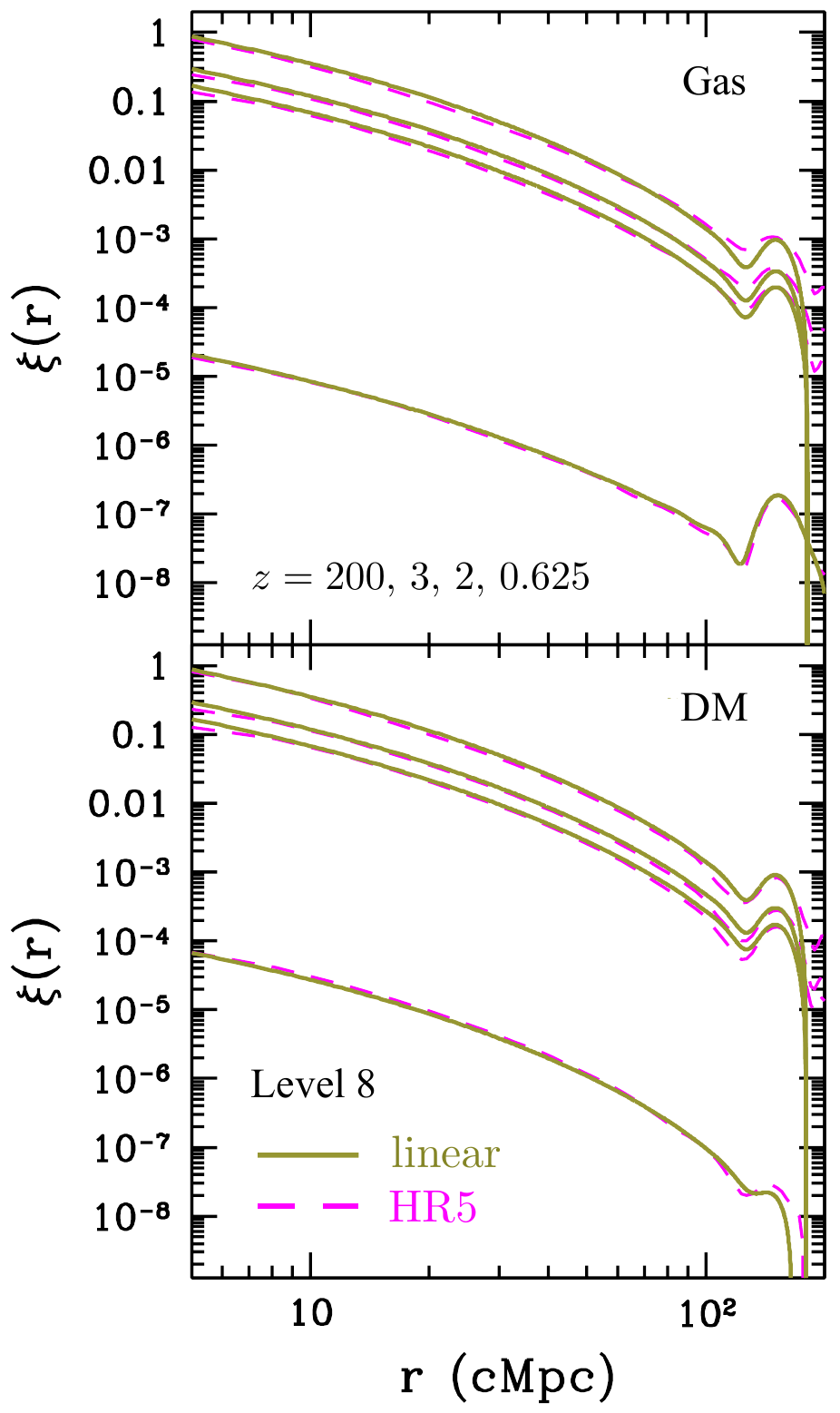}
    \caption{Matter correlation functions measured at the base level 8 at redshifts $z=200$, 3, 2, and 0.625. The measured correlation functions of dark matter (bottom) and gas (upper) are shown in the dashed magenta lines and the linear predictions are shown with the solid olive ones.}
    \label{fig:corr8}
\end{figure}

Figure \ref{fig:corr8} shows the measured correlation functions of the dark matter distribution (bottom panel) and the baryon distribution (top panel) at level 8 at redshifts $z=200$, 3, 2, and 0.625 (from bottom to top). These measurements are also made from the dark matter and gas density fields assigned on a $256^3$ voxel grid. The BAO feature in the baryonic correlation function is evident at $z=200$, but it gradually weakens with decreasing redshift while the dark matter clustering shows the opposite effect.
This is reflection of the gravitational coupling between baryon and DM over time, which is also seen in the power spectrum analysis given previously.

\subsection{Matter Clustering in the Zoomed Region} \label{sub:level13}

The two-point correlation functions in the zoomed region can directly be obtained by pairing two grid cells with distance $r$ as
\begin{equation}
    \xi(r) = {1\over V} \sum_{r^\prime} \delta(r^\prime)\delta(r^\prime+r),
\end{equation}
where $V$ is the total volume of the space of interest and $\delta(r)$ is the density contrast of the grid cells.
\begin{figure}
    \centering
    \includegraphics[width=1\columnwidth]{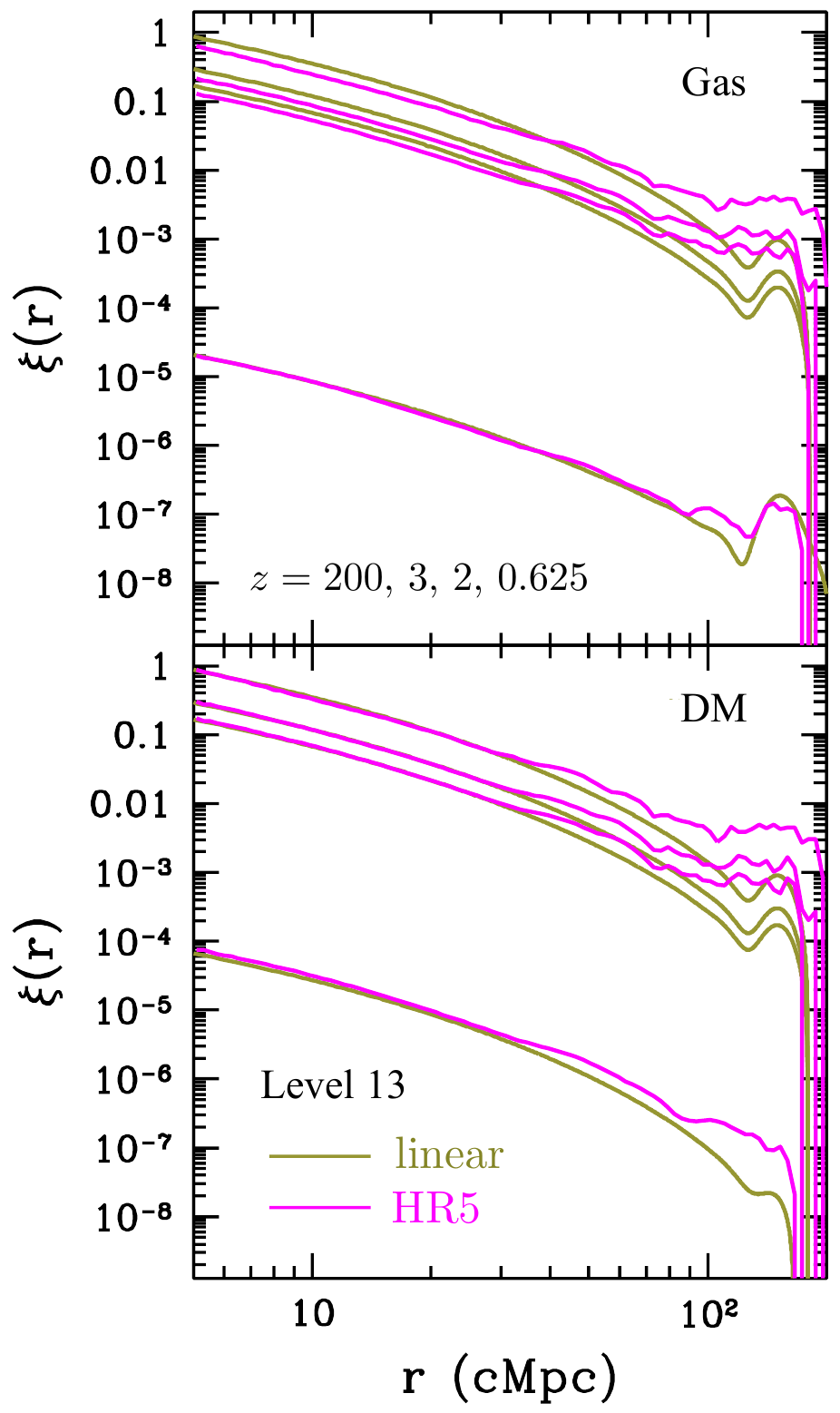}
    \caption{Same as Figure \ref{fig:corr8}, but measured in the zoomed region.}
    \label{fig:corr13}
\end{figure}
In the clustering measurements, we used only the grid cells left after applying the mask (see Appendix \ref{sec:geometry}) to minimize the boundary effects resulting from massive dark matter particles infiltrating from low level regions. Figure \ref{fig:corr13} shows the correlation functions of baryons (top) and dark matter (bottom) at redshifts $z=200$, 3, 2, and 0.625. At the initial redshift ($z=200$), dark matter clustering has an amplitude somewhat higher than expected in the vicinity of the BAO scale ($r\sim100-150\,\cMpc$) while baryons show the BAO peak that is well-matched with the linear theory. We note that baryons include both hot and cold gas components.

\subsection{Galaxy Clustering}\label{sub:structure}
To quantify the clustering strength of our simulated galaxies we calculate the two-point correlation function. We adopt the \citet{1993ApJ...412...64L} estimator, 
\begin{equation}
    \xi_{gg}(r) = { {\mathrm{DD}(r) - 2\mathrm{DR}(r) + \mathrm{RR}(r)}\over \mathrm{RR}(r)}, 
\end{equation}
where DD, RR, and DR are the numbers of galaxy-galaxy, random-random, and galaxy-random pairs, respectively.
The error bars are estimated using the bootstrap method recommended by \citet{1984MNRAS.210P..19B}. We use 100 samples for our bootstrap resampling, with the galaxies randomly re-sampled from the original catalogue. 

\begin{figure}
    \centering
    \includegraphics[width=1\columnwidth]{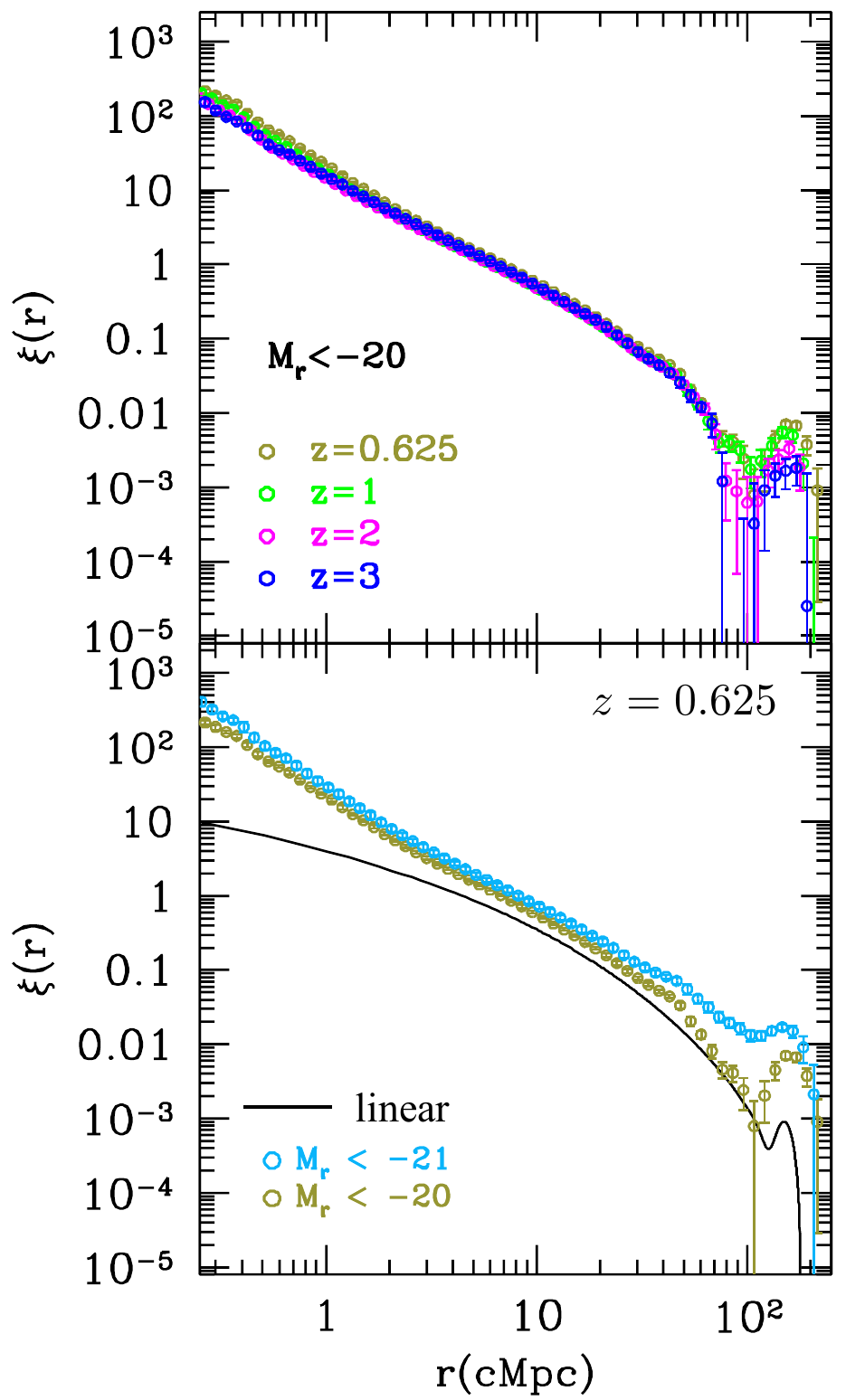}
    \caption{Galaxy correlation functions in the zoomed region. The bottom panel shows the galaxy clustering at $z=0.625$ for the samples of $M_{\rm r}<-21$ ({\it cyan}) and -22 ({\it olive color}). The black solid curve indicates the linear matter prediction and error bars are obtained from 100 bootstrap resamplings. In the top panel, the galaxy correlations in $M_{\rm r}<-20$ are displayed at the four different redshifts ($z=3$, 2, 1, and 0.625).}
    \label{fig:2pcf}
\end{figure}
In Figure~\ref{fig:2pcf}, we show the correlation function of the galaxies in \texttt{HR5} to inspect its redshift (top) and brightness (bottom panel) dependence. The linear correlation function at $z=0.625$ is also shown in the bottom panel for comparison. The relatively higher bias in the clustering of brighter galaxies is evident, and the effects of non-linear evolution is visible on both small ($r\lesssim5\,\cMpc$) and large ($r\gtrsim50\,\cMpc$) scales. 

It can be seen that the galaxy correlation function is well described by a single power-law function of a slope $\sim-1.65$ almost all the way down to at least $r\sim50\,\cMpc$ where it is no longer possible to calculate it accurately. On large scales, the BAO peak is detected in both magnitude bins. Astrophysics on small-scales seems to be responsible for the nonlinear bias in brighter galaxies even on very large scales, and is a classic example of the non-linear behaviour of galaxy clustering. It is also interesting to see the redshift evolution of clustering for the same absolute-magnitude samples as depicted in the top panel. With decreasing redshifts, the clustering amplitudes both on the small scale ($r\lesssim2\,\cMpc$) and on the BAO scale ($r\gtrsim100\,\cMpc$) show a substantial enhancement while the mid-scale clustering seems to be almost invariant over the period of the time.

Galaxy clustering is believed to be biased compared to the background dark matter distribution, since galaxy formation prefers high-density regions. Hence, more massive or brighter galaxies tend to form at higher density peaks and are therefore more biased. This brightness-dependent biasing can be seen in the bottom panel of the figure. The error bars represent the standard deviation obtained from  100 bootstrap re-sampled catalogs without correction \citep{2009MNRAS.396...19N}. When Figure~\ref{fig:2pcf} is compared with Figure~\ref{fig:corr13}, it can be noticed that the correlation function of galaxies can be very different from that of gas both in amplitude and shape. It should be pointed out that the baryon correlation function is measured by using all gas components. On the other hand, the galaxy correlation function represents the clustering of cold gas clumps.

It is noteworthy to compare the clustering of \texttt{HR5} and \texttt{TNG300} from the \texttt{IllustrisTNG} project \citep{Springel18}, which also addressed the BAO feature using the power spectrum and the bias. They showed the BAO features in the power spectrum by multiplying the ``linear'' BAO wiggle with the square of the bias measured using an empirical smooth function. However, the BAO wiggle is not observed directly from the power spectrum or two-point correlation of \texttt{TNG300} because the BAO feature is buried under Poisson noise. This is because the relatively small box size of \texttt{TNG300} means that it has a restricted number of independent Fourier modes around the BAO scale. On the other hand, the larger simulation volume of \texttt{HR5} enables us to examine the BAO peaks in the two-point correlations of galaxies, as seen in Figure~\ref{fig:2pcf}.

\begin{figure}
    \centering
    \includegraphics[width=1\columnwidth]{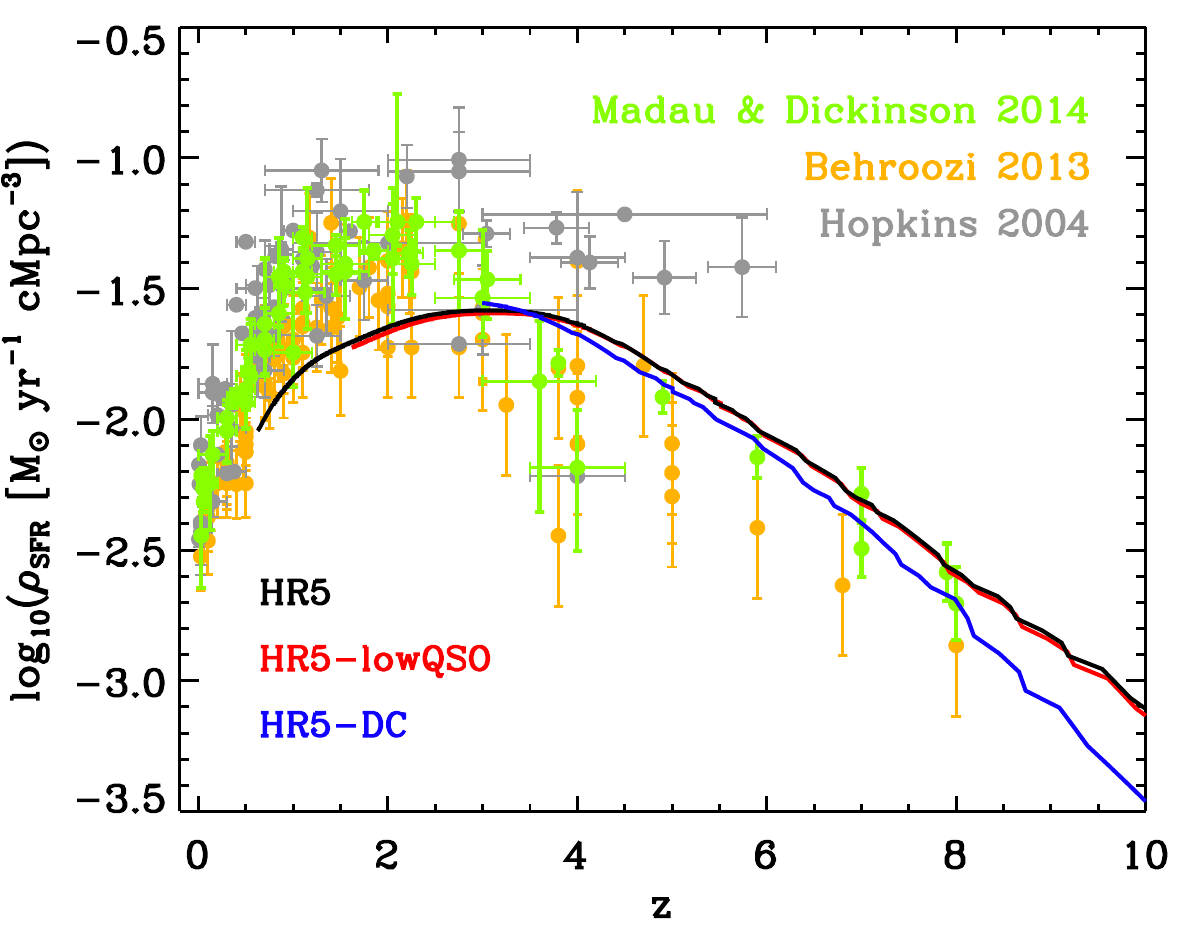}
    \caption{Cosmic star formation history until $z=0.625$, 1.5, 3.0 for \texttt{HR5} (black), \texttt{HR5-lowQSO} (red), and \texttt{HR5-DC} (blue), respectively. Green, grey and orange dots with error-bars represent the observed results from \citet{Hopkins04}, \citet{Madau_Dickinson14} and \citet{Behroozi13}.}
    \label{fig:sfh}
\end{figure}

\subsection{Cosmic Star Formation History}\label{sub:SF}

The CSFH measured in the three \texttt{HR5} runs is shown in Figure~\ref{fig:sfh}. As displayed in Table~\ref{tab:variables}, the stellar particles in \texttt{HR5} carry their birth epoch and initial mass information throughout the whole run, after their formation. The star formation rate density at time $t$ can be calculated using only the final snapshot, by summing the initial mass of the stellar particles that have birth epochs between $t$ and $t+\Delta t$. Since most stellar particles are initially born in galaxies ($\sim80$\% inside the self-bound objects identified by \texttt{PGalF}), we use not only the stellar particles bound to galaxies, but also those unbound at the final epoch because we assume that all stars formed in galaxies, and have been scattered via astrophysical and numerical effects. In order to alleviate potential boundary effects, we find stellar particles located in a volume $1\,l_{\rm corr}$ away from the surfaces contaminated by low level particles, where $l_{\rm corr}$ is the correlation length, and it is measured to be $4.09\,\cMpc$ at the final snapshot. The CSFH is only calculated using the stellar particles inside this non-contaminated volume. 

The key parameters which control the star formation history are the star formation efficiency, the stellar feedback and the AGN feedback. While the star formation efficiency most strongly affects the overall normalization of the star formation history, the AGN and stellar feedback impact the gradient of the different parts of the history. The early time star formation rate is also sensitive to the rare high density regions that the large volume of \texttt{HR5} simulation contains. Feedback from delayed cooling suppresses the star formation rate at early times, but causes it to subsequently increase around the peak star formation epoch. This is because the delayed cooling approach generates cold, dense and slow winds, which fall back onto the galaxies at later epochs.

The CSFH is the primary calibration point of the test runs for \texttt{HR5}, and thus, as expected, our overall CSFH compares well with the observations, particularly at high and low redshifts. The CSFH in \texttt{HR5} reaches a peak at $z\approx 3$, somewhat earlier than the observed peak at $z\approx 2$. Furthermore, the transition from increasing to decreasing seems more gradual in \texttt{HR5} than in the observations. This may be an artifact of the sub-grid physics or resolution. In Figure~8 of \citet{Kaviraj17}, the CSFH of \texttt{Horizon-AGN} is compared with the results of \citet{Hopkins06}, showing reasonable agreement. However, more recent analyses by \citet{Behroozi13} and \citet{Madau_Dickinson14} lowered the high redshift star formation history substantially (see their Figure~2). Our early time cosmic star formation is in better agreement with these data. It is also worth noting that the IMF plays a critical role in deriving the star formation rate density, $\rho_{\rm SFR}$, from observations. Among the three empirical studies cited in Figure~\ref{fig:sfh}, a Chabrier IMF is adopted in \citet{Behroozi13} and a Salpeter IMF \citep{salpeter55} is assumed in the rest. The estimate of $\rho_{\rm SFR}$ is 0.25\,dex lower with a Chabrier IMF than with a Salpeter IMF at a fixed luminosity~\citep{panter07}. This is because the Chabrier IMF is log-normal in $M<1\,\msun$, while the Salpeter IMF follows a simple power law in the whole mass range. This should be taken into consideration in the comparison between \texttt{HR5} and the empirical studies in Figure~\ref{fig:sfh}.

\subsection{Galaxy Stellar Mass Functions}\label{sub:SMF}

\begin{figure}
    \centering
    \includegraphics[width=1\columnwidth]{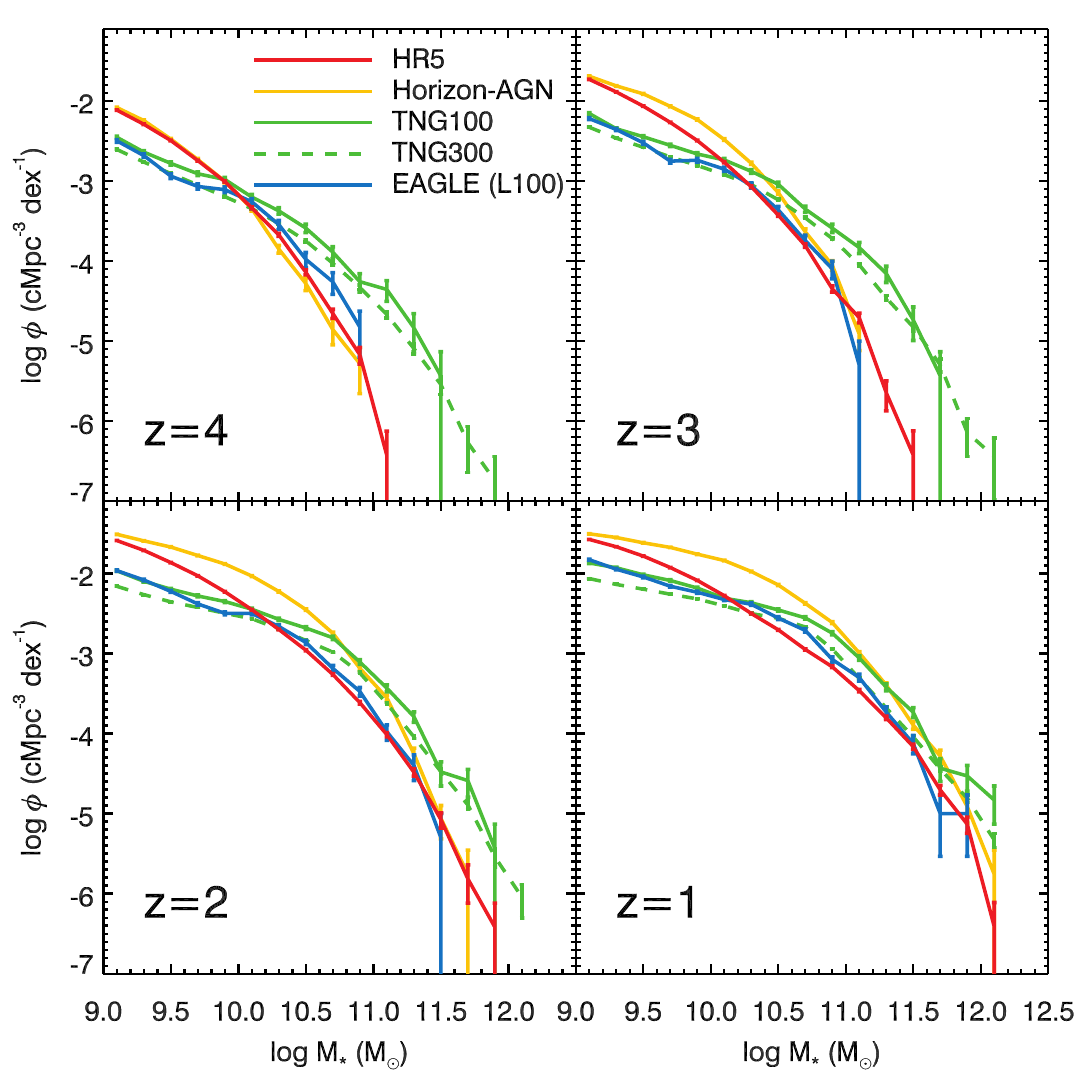}
    \caption{Galaxy stellar mass functions at $z=1-4$ in the zoomed region. The red, orange, green and blue  lines show the GSMFs of \texttt{HR5}, \texttt{Horizon-AGN} \citep{Dubois2014}, \texttt{TNG100} (solid) and \texttt{TNG300} (dashed) of the \texttt{IllustrisTNG} project 
    \citep{pillepich18b}, and \texttt{EAGLE (L100)} \citep{Schaye2015}, respectively. The error bars on the solid lines mark Poisson errors. 
    } \label{fig:gsmf}
\end{figure}

Galaxy stellar mass functions (GSMF) constitute one of the fundamental global properties that simulations should reproduce. Figure~\ref{fig:gsmf} shows GSMFs at $z=1-4$ in the zoomed region of \texttt{HR5} (red) along with previous cosmological hydrodynamical simulations, \texttt{Horizon-AGN} \citep[][orange]{Dubois2014}, \texttt{TNG100} and \texttt{TNG300} of the \texttt{IllustrisTNG} project 
\citep[][green]{pillepich18b}, and \texttt{EAGLE (L100)} \citep[][blue]{Schaye2015}, which have a spatial resolution for gas comparable with that of \texttt{HR5}, but in a smaller volume.  It is well known that the chosen galaxy and halo finding schemes do not produce significantly different results, in terms of the global properties of the DM and stellar components. However, this is not the case for the gas components because thermal energy is not taken into consideration in the unbinding process used by some halo finding schemes \citep{knebe11,knebe13}. On the other hand, \citet{pillepich18b} demonstrate that the massive end of the GSMF can be sensitive to the aperture size used for mass measurement, even using a given halo/galaxy finding scheme. In \texttt{TNG100}, \texttt{TNG300}, and \texttt{EAGLE (L100)}, galaxy stellar mass is defined to be the mass sum of stellar particles bound to subhalos identified by the \texttt{SUBFIND} algorithm~\citep{Springel2001,dolag09}. In \texttt{Horizon-AGN}, galaxies are found from stellar distribution using the \texttt{AdaptaHOP} algorithm~\citep{aubert04}. In all the simulations cited in Figure~\ref{fig:gsmf}, except \texttt{HR5}, galaxy stellar mass is that given by the substructure finding algorithms. In this study, as mentioned above, for the \texttt{HR5} curves, we measured galaxy stellar mass inside five times the half mass radius of stellar particles bound to individual galaxies. We note that the intrinsic differences induced by different galaxy finding and mass measurement schemes are not assessed here but will be discussed in a future paper.

The volume used to measure the GSMFs is the same as that used for the global star formation rates in Sect.~\ref{sub:SF}. 
\texttt{HR5} shows a GSMF the same as that of \texttt{Horizon-AGN} at $z=3-4$. However, in \texttt{HR5} galaxies grow more slowly than those in \texttt{Horizon-AGN} in $\log M_*/M_{\odot}\sim10-11$ at $z<3$. The GSMFs of \texttt{TNG100} 
and \texttt{EAGLE (L100)} are shallower than those of \texttt{HR5} and \texttt{Horizon-AGN} at all the redshifts in low mass range ($\log M_*/M_{\odot}<10$). This has been discussed in a couple of previous studies. \citet{Kaviraj17} claim that insufficient stellar feedback, and numerical resolution effects, may cause the steep slope at the low mass-end in the GSMF of \texttt{Horizon-AGN}. \citet{pillepich18b} demonstrate that an improved prescription for stellar-driven winds effectively suppresses the low mass-end in the GSMF in \texttt{TNG100}, compared to that of its predecessor, \texttt{Illustris}. \texttt{EAGLE (L100)} happens to agree with \texttt{HR5} above the mass range at all the redshifts. Interestingly, with decreasing redshift, all the simulations converge to each other at the massive end. In a follow-up study, we are testing observational techniques to measure mass and luminosities from the mock galaxy images of \texttt{HR5}. This will help understand differences between the GSMFs measured from simulations and observations.

\begin{figure}
    \centering
    \includegraphics[width=1\columnwidth]{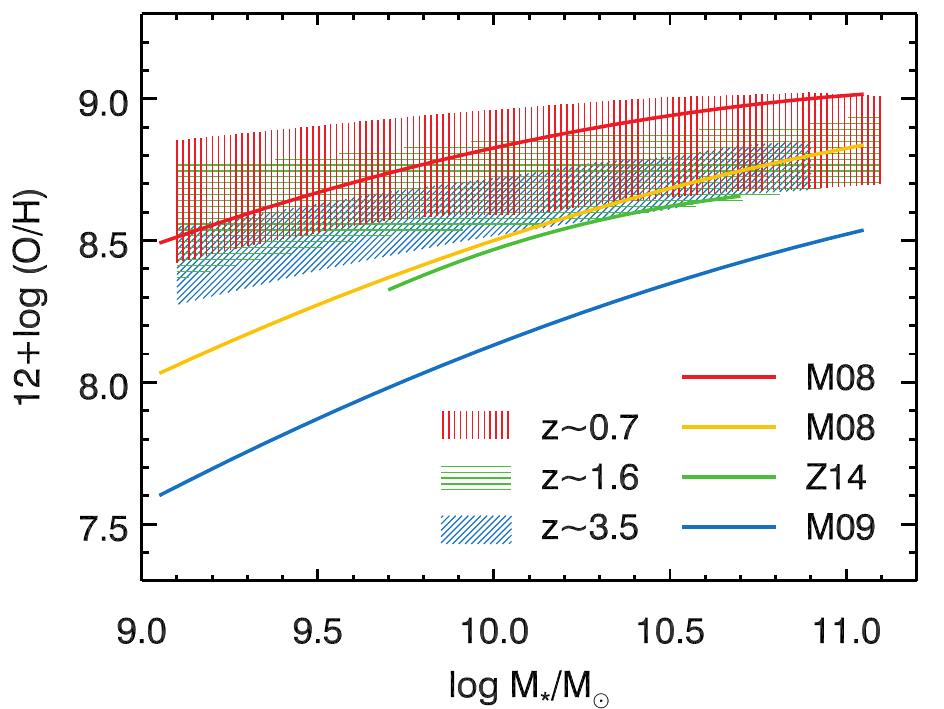}
    \caption{Gas phase metallicity ($12+\log (\rm O/H)$) versus galaxy stellar mass relation for \texttt{HR5} and observations at $z\sim0.7-3.5$. The solid lines denote empirical fits presented in ~\citet[][M08]{maiolino08} at $z\sim0.7$ (red) and $z\sim2$ (orange), \citet[][Z14]{zahid14} at $z\sim1.6$ (green), and \citet[][M09]{mannucci09} at $z\sim3.5$ (blue). The hatched regions mark the $2.5^{\rm th}$ to $97.5^{\rm th}$ percentile distributions of the gas phase metallicities of the \texttt{HR5} at the redshifts, similar to those of the empirical data.
    } \label{fig:chem}
\end{figure}

\subsection{Gas Phase Metallicity}\label{sub:chem}
The gas phase metallicity for the \texttt{HR5} galaxies $\rm 12+\log(O/H)$ is measured within a fixed aperture. Figure~\ref{fig:chem} shows the gas phase metallicity versus galaxy stellar mass relation of \texttt{HR5} and for multiple observational data set between $z\sim0.7-3.5$. The empirical relation denoted by the red and orange solid lines come from the fitting formula derived by \citet{maiolino08} from galaxies observed at $z\sim0.7$ \citep{savaglio05} and $z\sim2$ \citep{erb06}. The green solid line indicates the empirical fit presented in \citet{zahid14} for galaxies observed by the FMOS-COSMOS survey at $z\sim1.6$. The blue solid line is the fit for the AMAZE and LSD galaxies found to lie between $z\sim3-4$~\citep{maiolino08,mannucci09}. The hatched regions show the $2.5^{\rm th}$ to $97.5^{\rm th}$ percentile distributions of the gas phase metallicity of the \texttt{HR5} galaxies. Since the gas phase metallicity is sensitive to aperture size, we adopt an aperture $7\,\kpc$ in diameter. This size is used to aid easy comparisons with the observations of the high redshift spectroscopic observations of \citet{savaglio05}, \citet{erb06}, and \citet{maiolino08}. Meanwhile, the FMOS-COSMOS survey utilized in \citet{zahid14} adopts fibers with the aperture size of 1.2\,arcsec, which corresponds to $\sim10\,\kpc$ at $z\sim1.6$. Therefore, even if the observations cited above are aperture-corrected, the empirical data should be carefully compared with each other. One can see differences between simulated galaxies and observations, particularly at low mass. Galaxies in \texttt{HR5} are too metal-rich at $z>2$, but somewhat metal poor at the more massive end at $z<1$. This is closely connected to the weak downsizing trend of of \texttt{HR5} hinted at in Figure~\ref{fig:gsmf}. In \texttt{HR5}, like in \texttt{Horizon-AGN}, small galaxies are formed and grow more rapidly than the downsizing trend suggests. Accordingly, the galaxies in \texttt{HR5} also suffer from overly rapid early chemical evolution at the low mass end, resulting in the shallow slopes shown in Figure \ref{fig:chem}. As discussed in Sect.~\ref{sub:SMF}, this can be attributed to incomplete stellar feedback and resolution effects. Indeed, a steeper mass-metallicity relation is seen in \texttt{IllustrisTNG}, which shows GSMFs flatter than \texttt{Horizon-AGN} and \texttt{HR5} at low mass end with effective stellar winds~\citep{torrey19}. Furthermore, the gas phase metallicity is steeper in higher resolution simulations such as {\tt MaGICC} \citep{obreja14}, and {\tt NewHorizon}, a $40\,{\rm pc}$ zoom-in resimulation of \texttt{Horizon-AGN} \citep{dubois20}. Outflows typically have metallicities well above the ISM average, suggesting that metals are ejected from galaxies before they have had a chance to mix with the ISM \citep{stinson12}. A fraction of these metals may be re-accreted at a later stage, but in any case will have a much longer mixing time with the ISM \citep{brook14}. However, such a process is very difficult to capture with kpc-scale resolution galaxy formation simulations.

\begin{figure}
    \centering
    \includegraphics[width=0.9\columnwidth]{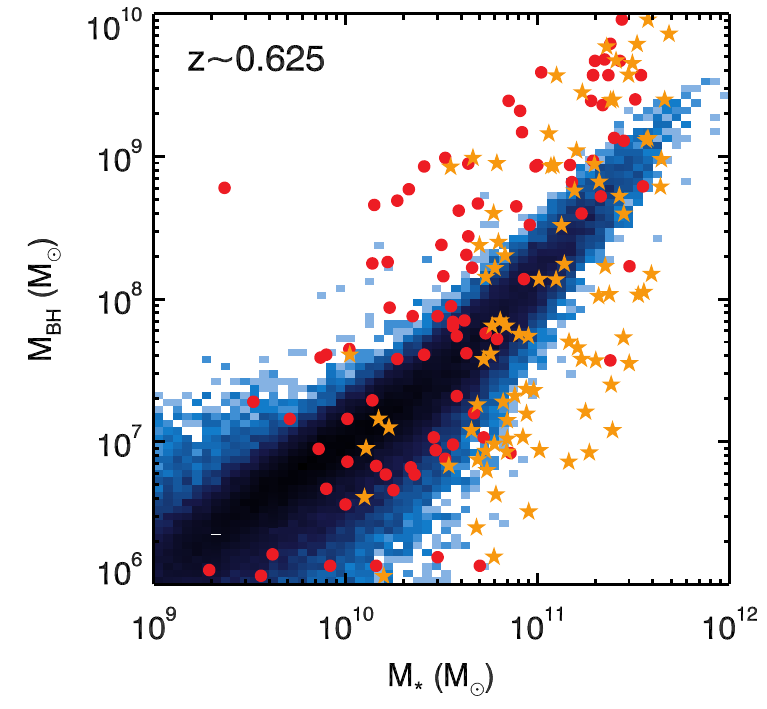}
    \caption{SMBH-host galaxy stellar mass relation for \texttt{HR5} at $z=0.625$ (blue shade) and local observations. Red dots and orange stars show empirical data from \citet{Reines15} and \citet{terrazas17}, respectively. The darker the shade is, the more galaxies there are. At $z=0.625$, the relation is shallower than the observed local relation, indicating that massive SMBHs in \texttt{HR5} grow less rapidly than needed to match the empirical relation.
    } \label{fig:bhsm}
\end{figure}

\begin{figure*}
    \centering
        \includegraphics[width=\textwidth, angle=0,scale=1.]{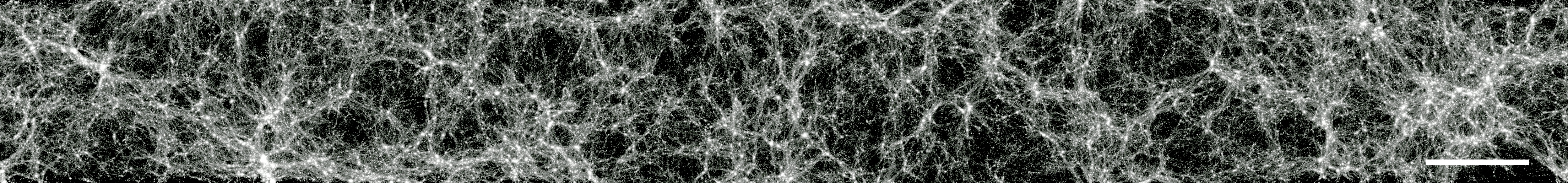}
    \includegraphics[width=\textwidth, angle=0,scale=1.]{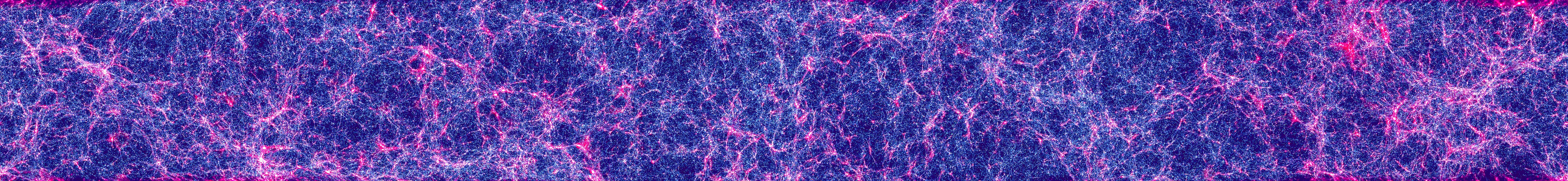}
    \includegraphics[width=\textwidth, angle=0,scale=1.]{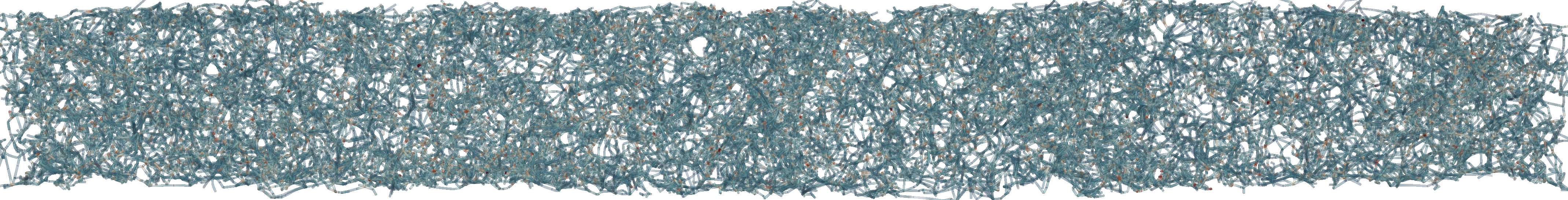}

    \caption{Full views of DM (top), gas (middle), and skeleton  structures (bottom) in the high resolution region in \texttt{HR5}. The skeleton structure is extracted from the galaxy distribution at $z=2$. In the top panel, DM particles are  more numerous in brighter regions. The reddish and bluish color code in the middle panel indicate gas temperature and density, respectively. The white horizontal bar in the top panel represents a scale length of $100\,\cMpc$.
    } \label{fig:HRskel}
\end{figure*}

\subsection{BH Formation History}\label{sub:BH}
By comparing BHs masses to properties of their host galaxies, we can probe the complex interaction between star formation, SN feedback, BH growth and AGN feedback which shape the properties of the galaxy population. Figure~\ref{fig:bhsm} shows the total BH-galaxy mass distribution from \texttt{HR5} at $z=0.625$ (the blue region) and the empirical distribution in the local universe from~\citet{Reines15} and~\citet{terrazas17}. Overall, the simulated galaxies at $z=0.625$ seem to be situated on the local empirical relation. At the low-mass end, the simulated and observed masses of BHs and galaxies are well matched, but BH masses in mid- and high-mass galaxies are underestimated. In comparison with \texttt{Horizon-AGN}, \texttt{HR5} has BHs $\sim0.5\,$dex less massive at given galaxy mass \citep[See Figure 12 in][]{Volonteri16}. As described in Sect.\ref{sub:feedback}, the physical ingredients for AGN feedback have been updated since the release of \texttt{Horizon-AGN} in \texttt{RAMSES}. The new prescription with the spins of BHs, coupled with QSO mode feedback, slows the growth of BHs down at high redshifts. This may be caused by the strong QSO mode feedback in our standard model, coupled with the spins of BHs, which slows down the growth of BHs at high redshifts. The BHs eventually catch up with the local scaling relation in low mass galaxies before $z=0.625$, but they are still below it in massive galaxies. Meanwhile, there have been many efforts to constrain the evolutionary trend of the $M_{\rm BH}-M_{*}$ relation over cosmic time. Some claim that BHs grow more efficiently than their host galaxies at high redshift \citep[e.g.][]{peng06,treu07,woo08,merloni10,decarli10,park15,caplar18}, while the others support no or weak evolution of the scaling relation \citep[e.g.][]{shields03,salviander07,shen08,cisternas11,shen15,sun15,suh20}. The formation and co-evolution of SMBHs and their host galaxies will be investigated in depth in future studies.

\subsection{Large scale structure mapping}\label{sub:cosmicweb}

\begin{figure*}
    \centering
    \includegraphics[width=0.8\textwidth]{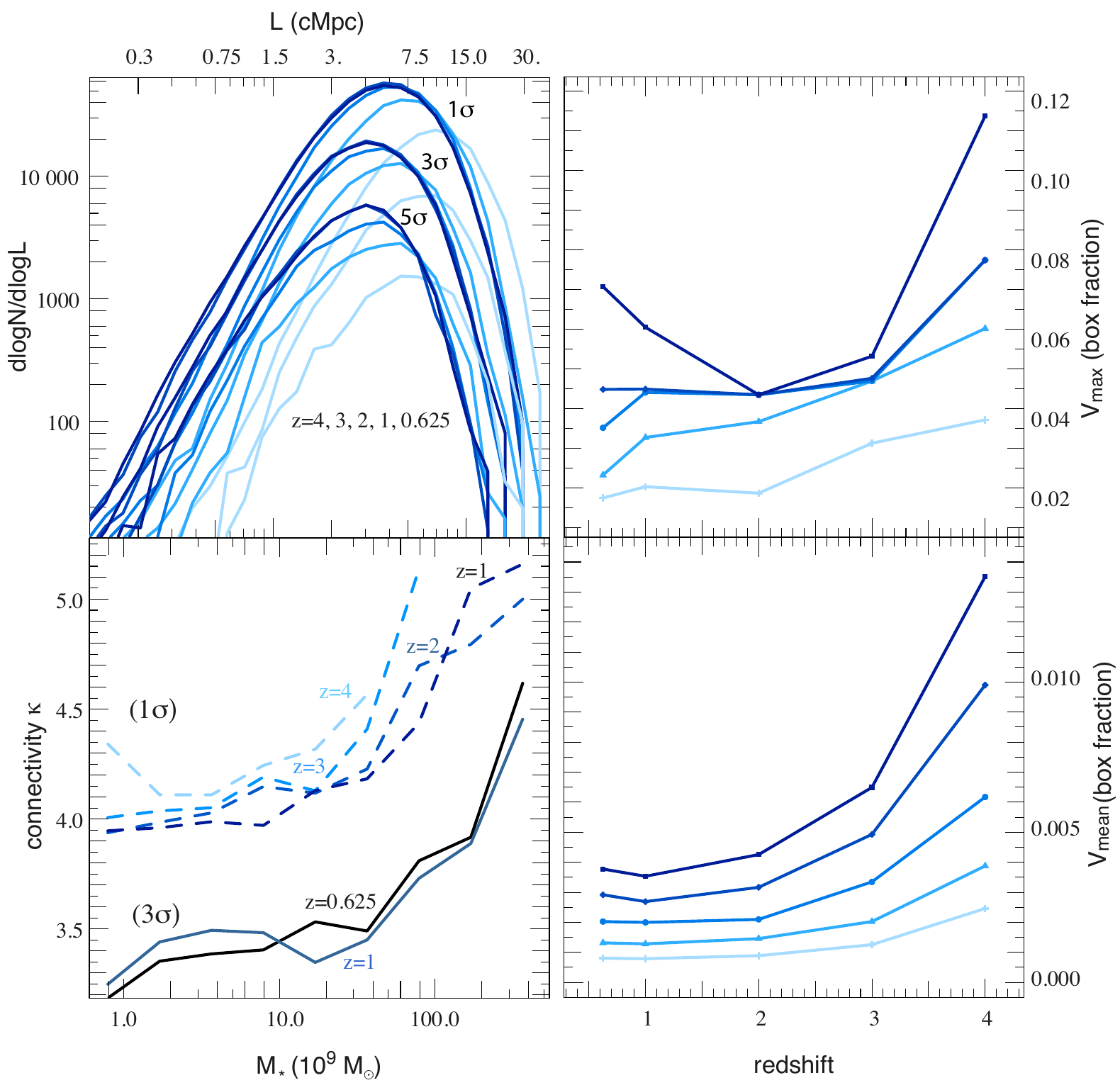}

    \caption{Top left: Length of filaments as a function of redshift (from light, high redshift, to dark, low redshift) and persistence in  \texttt{HR5} as labeled. The typical filament length is of the order of $5-10\,\cMpc$. Bottom left: Connectivity as a function of stellar mass and redshift 
 for 1 and $3\sigma$ persistence   in  \texttt{HR5} as labeled.
 Note that connectivity increases with stellar mass and redshift, as expected. Right panels: Cosmic evolution of relative (maximum, top, mean, bottom) void size for persistence level of 1 to $5\sigma$  (from  light to dark)
    } \label{fig:HRskeleton}
\end{figure*}

To trace the filaments of the cosmic web, we use the 3D ridge extractor {\tt DisPerSE}  \citep{sousbie112},  which identifies the so-called skeleton (gradient lines connecting peaks together through saddle points) as  1D-ascending manifolds of the discrete Morse-Smale complex \citep{Forman2002}. This complex is defined by the tessellation of the galaxy distribution. This scale-free algorithm relies on topology  to identify robust components of the cosmic web network. The result is quantified in terms of significance  compared to a discrete random Poisson distribution, through a quantity known as persistence (i.e. the ratio of the density at the peak and saddle point connected by a given ridge).  From the positions in the  galaxy catalogue alone, we extract 3$\times$5 skeletons corresponding to persistence levels of 1, 3 and 5$\sigma$ at redshifts 0.625, 1, 2, 3 and 4, respectively.  We purposely ignore galaxy stellar masses, since accurate estimates might not be available in deep surveys. In practice, however, due to the resolution limit, galaxies more massive than $10^9\,\msun$ are predominantly used in extracting the cosmic web from \texttt{HR5}.

Figure~\ref{fig:HRskel} shows the matter distribution, and the skeleton, of the entire high resolution region in the $z=2$ snapshot. From this skeleton we can characterise the typical distribution of filament length, shown in the top left panel of Figure~\ref{fig:HRskeleton},  to emphasise that we do measure long filaments within the \texttt{HR5} box. The bottom left of Figure~\ref{fig:HRskeleton} displays the redshift evolution of the connectivity as a function of stellar mass and redshift, for persistence values of one and three $\sigma$ (the $5\sigma$ curve is close to  $3\sigma$ one, and is not shown for clarity). The connectivity, $\kappa$, is defined as the number of filaments connected from a given node of the skeleton to its persistent saddle points. The high statistics of \texttt{HR5} allow us to extend  the dark matter halo trend found in \cite{codis2018} to  galaxies:  more massive galaxies are more connected, while the connectivity decreases with cosmic time. It also displays evidence of an elbow near $5\times10^{10}\,\Msun$, sightly increasing with cosmic time, following the  mass of non-linearity at that redshift (up to the baryonic abundance ratio). Rarer, more massive galaxies sit within  more isotropic, multiply connected, environments. This is expected given the impact of gravitational clustering in disconnecting filaments, \citep[see][for  theoretical motivation]{cadiou2020}. It is also consistent with the trend found at redshift zero in \citet{Kraljic:2019tz} from the hydrodynamical simulations {\tt Horizon-AGN}, {\tt Simba} and the {\tt SDSS} survey.  Finally, the top and bottom right panels of Figure~\ref{fig:HRskeleton} shows the largest and mean void size --- defined here as an ascending 3-manifold within {\tt DisPerSE} ---  as a function of redshift for persistence one to five $\sigma$ as labeled. One of the assets of  \texttt{HR5} is to probe  ($5\sigma$) persistent voids as large as spheres of radius $\sim 100\,\cMpc$ within the simulation.

\section{Summary and discussion}  \label{sec:conclusion}

\texttt{HR5} is the first cosmological to simulation to recover cosmic structures on Gpc scales, while also following the non-linear baryonic physics of galaxy formation down to $1\,\kpc$.  Starting from an initial redshift of $z=200$, the simulation reaches  $z=0.625$. Within a cubic simulation volume of $(1049~\mathrm{cMpc})^3$, we defined a cuboid-shaped region of $1049\times119\times127\,\cMpc^3$ where we carried out high resolution calculations (minimal spatial scale of 1 physical kpc). This zoomed volume is up to 10 times larger than those of the previous cosmological hydrodynamical simulations with a comparable spatial resolution (e.g. \texttt{TNG100} of the \texttt{IllustrisTNG} project, \texttt{Eagle (L100)}, and \texttt{Horizon-AGN}). It is higher resolution than \texttt{TNG300}, but has half the usable volume, despite capturing larger scales. 
Along with the snapshots, we generate additional data products. Light cone data was generated on-the-fly and additional data from five dense regions surrounding massive clusters was captured with a very high output cadence. By $z=0.625$ the high-resolution region contains 290,086 galaxies with stellar masses $M_{*}>10^9\,\Msun$, and a total of 102 virialized halos with total mass $>10^{14}\,\Msun$. The most massive halo in the simulation was found to have total mass of $M_{\rm tot}=5.2\times10^{14}\,{\rm M_\odot}$. The large volume of \texttt{HR5} enables us to investigate cosmological structures in a very wide range of scales, from filaments of a few cMpc length to voids as large as spheres of radius $\sim100\,\cMpc$.

On top of radiative cooling, star formation and stellar feedback, \texttt{HR5} adopts a dual jet-heating mode in the AGN feedback, and also includes chemical evolution, tracing the relative abundance of oxygen and iron. 

Comparisons of \texttt{HR5} with the observed global properties of the Universe and previous numerical studies, highlight both reasonable agreement and some discrepancies. The cosmic star formation history in \texttt{HR5} is reasonably consistent with  existing observations, even though some differences in the evolution are noticeable around the cosmic noon. The metallicity of the gas phase in simulated galaxies,  while in a similar range of values than observations at the massive end, shows that less massive galaxies are too metal-rich. 
This is a well known effect of a combination of weak stellar feedback and inadequate metal mixing leading to a scenario where the simulation produces a realistic amount of metals, but distributes them incorrectly
in the simulation volume.

Because the volume of \texttt{HR5} is large enough to contain very long wavelength modes in the initial power spectrum up to 1049~cMpc, it is able to track non-linear structure formation processes up to the BAO scale. This  can be seen in the two-point correlation function of massive galaxies ($M_{*}>10^{10}\,{\rm M_\odot}$). However, the correlation function on large ($\gsim 60\,\cMpc$) scales is much smaller than in linear theory, and is even negative for low-mass galaxies. This may indicate that hydrodynamic effects on small-scales can modify galaxy clustering on scales comparable to that of the BAO. Therefore, the use of the BAO as a cosmic standard ruler should be carefully examined to understand this tracer dependence. 

\texttt{HR5}'s largest void, as identified by {\tt DisPerSE}, demonstrates the unique ability of this large volume simulation to study the impact of large voids (up to $100\,\cMpc$) on galaxy formation. It allows us to reliably quantify the cosmic evolution of galactic connectivity, while the low-redshift measurements of mass dependency on the large-scale structure are found to be in fair agreement with both the {\tt SDSS} and other cosmological simulations. 

The \texttt{HR5} run will soon be used to address key challenges of numerical galaxy and cluster physics. In the short term, we plan to explore the cosmic evolution of the luminosity of high redshift AGNs (in optical X ray and radio wavelengths), the distribution of Ly-$\alpha$ emitters,  the geometry of shock waves around protoclusters, and the distribution of diffuse stellar components in the IGM. We also intend to explore, self-consistently, strong lensing around the clusters. The simulation will also be used to predict the impact of baryons (noticeably in clusters) on the shape of the inferred dark matter power-spectrum and the distribution of voids. More generally, the impact of feedback on the topology of large-scale structures will be investigated. Finally, at the technical level, thanks to its larger statistical sample and volume, \texttt{HR5} will be used to train deep-learning tools over less biased samples of galaxies.

As a byproduct of producing, and post-processing, this set of simulations,  we developed a new  hybrid \texttt{OpenMP-MPI} parallelization scheme for \texttt{RAMSES} to take advantage of the  many-core many-thread {\tt Nurion} Supercomputer. We also extended the standard Friends-of-Friend algorithm, and developed a new galaxy finder \texttt{PGalF} to analyse the large outputs of {\tt HR5}.  The corresponding algorithms together with the catalogues presented in this paper will be made publicly available. In the meantime please contact the first author for specific requests.


\acknowledgments
We thank the anonymous referee for constructive comments and a careful reading of the manuscript.
ONS acknowledges funding from \href{http://www.dimacav-plus.fr}{DIM ACAV+}.
BKG and CGF acknowledge the UK's Science \& Technology Facilities Council (STFC) through the University of Hull Consolidated Grant  ST/R000840/1.
SEH was supported by Basic Science Research Program through the National Research Foundation of Korea funded by the Ministry of Education (2018\-R1\-A6\-A1\-A06\-024\-977).
CNCP was partially supported by the Spin(e) grant ANR-14- BS05-0005 (\href{http://cosmicorigin.org/}{cosmicorigin.org}) and the Segal grant ANR-19-CE31- 0017  (\href{http://secular-evolution.org/}{secular-evolution.org}) of the French Agence Nationale de la Recherche. 
JK was supported by a KIAS Individual Grant (KG039603) via the Center for Advanced Computation at Korea Institute for Advanced Study.
The research of JD is supported by STFC and the Beecroft Trust.
JL thanks the TNG collaboration and the Virgo Consortium for granting him access to the data used in this paper.
This work benefited from the outstanding support provided by the KISTI National Supercomputing Center and its Nurion Supercomputer through the Grand Challenge Program (KSC-2018-CHA-0003). Large data transfer was supported by KREONET, which is managed and operated by KISTI.
We acknowledge ongoing access to {\sc viper}, the University of Hull High Performance Computing Facility. 
We thank St\'ephane Rouberol for the smooth running of the HORIZON Cluster, where some of the post-processing was carried out.
We thank Thierry Sousbie for provision of the {\texttt{DisPerSE}} 
code (\href{http://ascl.net/1302.015}{ascl.net/1302.015}). 
CNCP thanks Sandrine Codis for advice.


\software{\texttt{RAMSES} \citep{Teyssier02}, 
		\texttt{CAMB} \citep{lew00}, 
		\texttt{MUSIC} \citep{hah11}, 
		\texttt{DisPerSE} \citep{sousbie112}}


\appendix

\section{Simulation Parameter Tuning}\label{sec:simparamtuning}

The primary calibration points of test runs for \texttt{HR5} are the CSFH, the GSMF, and the mass-metallicity relation. Table~\ref{tab:hr5p} shows the list of parameters and their ranges tuned for model calibration. In the test runs, we searched for a parameter set producing the CSFH and GSMF trends that agree with observations within an uncertanty of $3\sigma$ . In the case of stellar mass--metallicity relation, we could not find any parameter set which reproduced the relation across the whole stellar mass range, (as discussed in Sect. \ref{sub:chem}). Therefore, we aimed to match the overall metallicity, instead of accurately following the observations in detail. We note that this calibration was evaluated by eyes, with no formal fitting, due to limited test resources.

The star formation efficiency directly controls galaxy growth and chemical enrichment of galaxies in the whole mass range. The mode of SN feedback and its efficiency significantly affect the number density of small galaxies, chemical enrichment of whole galaxies, and the slope of the CSFH at high redshifts. AGN feedback mainly regulates the growth of massive galaxies and the slope of the CSFH at low redshifts.

\begin{deluxetable}{ccccccccc}[hbt]
\tablehead{
\colhead{(a) Name} & 
\colhead{(b) $L_{\text{box}}$ (cMpc)}& 
\colhead{(d) $\epsilon_*$} & 
\colhead{(d) SN} & 
\colhead{(e) $\eta_\text{SN}$} &
\colhead{(f) $f_w$} &
\colhead{(g) $\epsilon_f$} & 
\colhead{(h) Delayed Cooling} &
\colhead{(i) Main Run}
}
\caption{Parameter tuning performed with different sub-grid galactic models, parameters for the SN and AGN feedbacks, and box sizes.  (a) Name of test run. (b) Size of the simulation box in a comoving scale. (c) star formation efficiency. (d) Presence of feedback from SNe: ``T'' stands for the thermal feedback and ``K'' stands for the kinetic feedback with $\epsilon_\text{K}=0.3$ of total released SN energy. (e) SN energy release ($\times 10^{51}\,{\rm erg}$). (f) Mass loading factor of the jet from SNe. (g) AGN feedback efficiency (radio/QSO modes). (h) AGN energy delay. AGN feedback includes both the radio and quasar mode. (i) The name of main run. }
\label{tab:hr5p}
\startdata
1 & 16.4 & 0.03 &T & 1 & 0  & 1/0.15 & No & \\
2 & 16.4 & 0.03 &T & 1 & 0  & 1/0.15 & Yes &\\
3 & 16.4 & 0.03 &T & 1 & 0  & 1/0.015 & No &\\
4 & 16.4 & 0.03 &T & 1 & 0  & 1/0.01 & Yes &\\
5 & 16.4 & 0.03 &T+K & 1 & 1  & 1/0.15 & No &\\
6 & 16.4 & 0.03 &T+K & 1 & 3  & 1/0.15 & No &\\
7 & 16.4 & 0.03 &T+K & 1 & 3  & 1/0.15 & Yes &\\
8 & 16.4 & 0.03 &T+K & 2 & 3  & 1/0.15 & No &\\
9 & 16.4 & 0.03 &T+K & 2 & 3  & 1/0.15 & Yes &\\
10 & 16.4 & 0.03 &T+K & 1 & 3  & 1/0.015 & No &\\
11 & 16.4 & 0.02 &T+K & 1 & 10  & 1/0.15 & No &\\
12 & 32.7 & 0.02 &T+K & 2 & 1  & 1/0.15 & No &\\
13 & 32.7 & 0.02 &T+K & 2 & 1  & 1/0.15 & Yes &\\
14 & 32.7 & 0.02 &T+K & 2 & 2  & 1/0.15 & No &\\
15 & 32.7 & 0.02 &T+K & 2 & 2  & 1/0.15 & Yes &\\
16 & 32.7 & 0.02 &T+K & 2 & 5  & 1/0.15 & No &\\
17 & 32.7 & 0.02 &T+K & 2 & 10  & 1/0.15 & No &\\
\hline
18 & 16.4 & 0.02 & T+K & 2 & 3  & 1/0.015 & No & \texttt{HR5-lowQSO}\\
19 & 32.7 & 0.02 & T+K & 2 & 3  & 1/0.15 & Yes & \texttt{HR5-DC}\\
20 & 32.7 & 0.02 & T+K & 2 & 3  & 1/0.15 & No &  \texttt{HR5}\\
\enddata
\end{deluxetable}

\newpage


\section{Geometry evolution of the zoomed region}\label{sec:geometry}
\begin{figure}
    \centering
        \includegraphics[width=\textwidth, angle=0,scale=0.8]{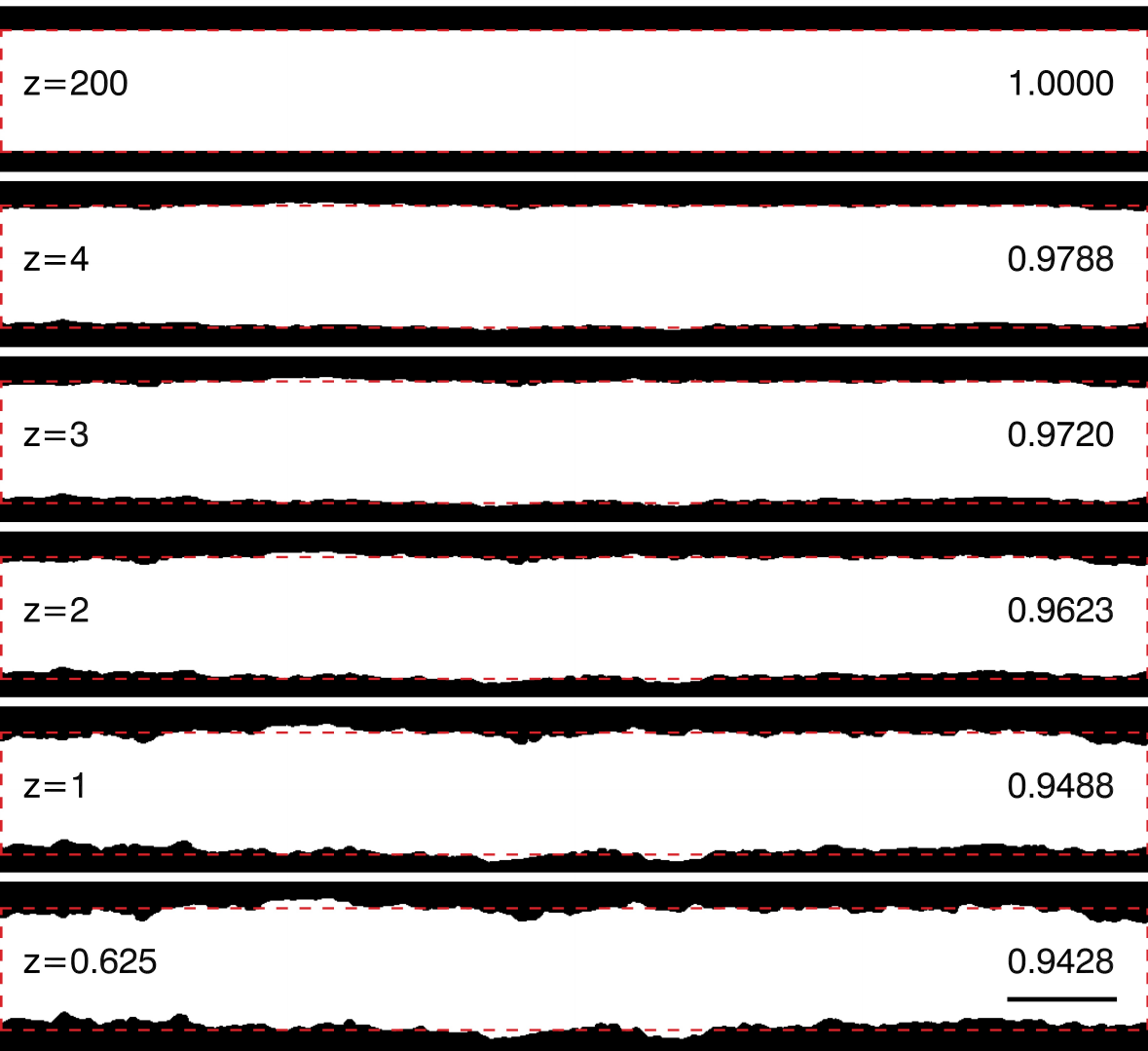}

    \caption{Projections of the region occupied only by the highest level particles (the zoomed region, white areas) and its neighbouring region filled or contaminated by low level, high mass particles (outside the zoomed region, black areas) at $z=200$ (IC), 4, 3, 2, 1, and 0.625. The slices projected in this figure have $\pm10\,\cMpc$ of thickness from the central axis, for illustration. The red dashed boxes mark the initial zoomed region at $z=200$. The number at the far right indicates the fraction of the volume at each redshift to the initial zoomed volume. The black horizontal bar in the bottom right denotes the scale of $100\,\cMpc$.
    } \label{fig:zoom_geometry}
\end{figure}

The unique geometry of the zoomed region in \texttt{HR5} is a compromise between the necessity of constructing light cone space data, $\sim1\,{\rm cGpc}^3$ scale volume, up to $\sim 1\,\kpc$ of spatial resolution, and computational resource constraints. The boundary of the high resolution region is initially made of flat surfaces, but they become bumpy as time passes due to the evolution of the density field. Thus, the zoomed region is inevitably contaminated by low level particles around the boundary. Figure~\ref{fig:zoom_geometry} shows the geometry of the uncontaminated region (white) at $z=200$ (IC), 4, 3, 2, 1, and 0.625. The boundary between the black and white areas has shrunk inward when a part of the zoomed region is denser than a neighbouring low level region. The volume of the uncontaminated region gradually decreases with decreasing redshifts as low level particles permeate into the zoomed region. The distance to the boundary is also provided in the galaxy catalogue of \texttt{HR5}, for those who may need to select galaxies barely affected by the low level region.

\section{Consistency Check for the Zoomed Region}\label{sec:consistency}
\begin{figure}
    \centering
        \includegraphics[width=0.8\textwidth]{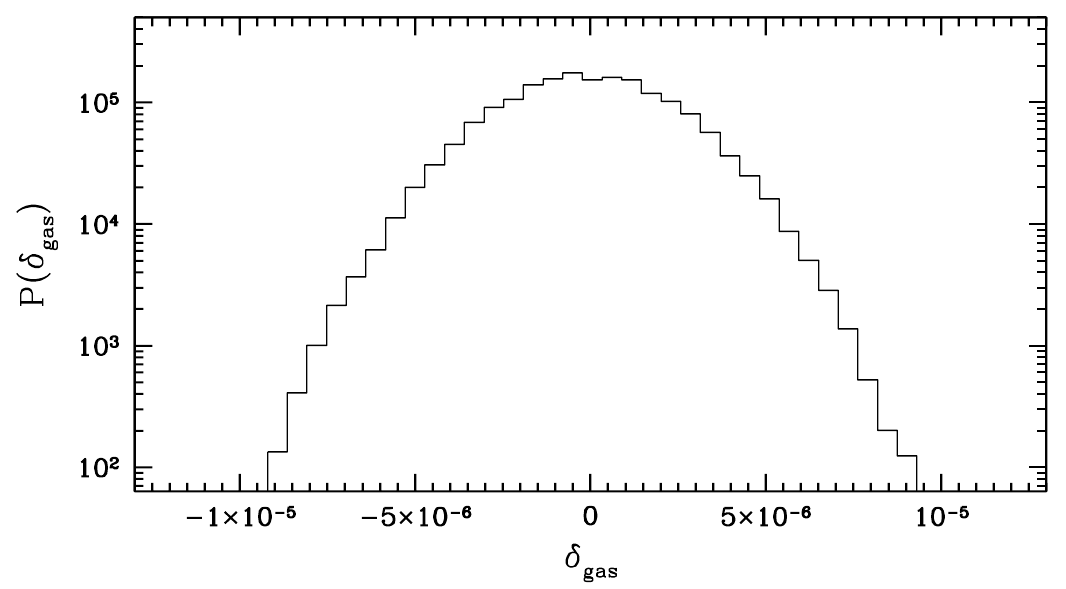}
    \caption{Probability distribution of the density differences between the base-level grids (level 8) and the grids of the same dimension reconstructed from the cells at level 13 in the initial conditions ($z=200$).
    } \label{fig:zoom_consistency}
\end{figure}

We conduct a test to verify the consistency of the initial conditions in the zoomed region, where we employed the stitching scheme to construct the elongated cuboid geometry. All the IC patches presented in Sect.~\ref{sub:IC} have the same matter distribution at the base-level, and they are stitched together at the levels higher than the base-level. If the stitching is credible, the density field at the base-level should be identical to the density field reconstructed from the matter distribution at the highest level. Since the base-level of \texttt{HR5} is composed of $256^3$ grids, we measure the density field at the highest level at the same number of grids, and examine the difference between the two density fields as follows:
\begin{equation}
    \delta (\mathbf{x}) \equiv {\rho_{13}(\mathbf{x})-\rho_{8}(\mathbf{x}) \over \rho_8(\mathbf{x})},
\end{equation}
where $\rho_{13}$ is the density field constructed from the cells at level 13, i.e. the zoomed region of \texttt{HR5} and the highest level at $z=200$, and $\rho_8$ is the density field at the base-level.
Figure \ref{fig:zoom_consistency} shows the histogram of $\delta_{\rm gas}$. Most of the grids have density differences $|\delta_{\rm gas}|$ smaller than about $10^{-5}$, which may be partly caused by the single precision accuracy adopted in this analysis. From this figure, we may conclude that the density field in the zoomed region well follows the density field at the base level.

\begin{figure}
    \centering
        \includegraphics[width=0.85\textwidth]{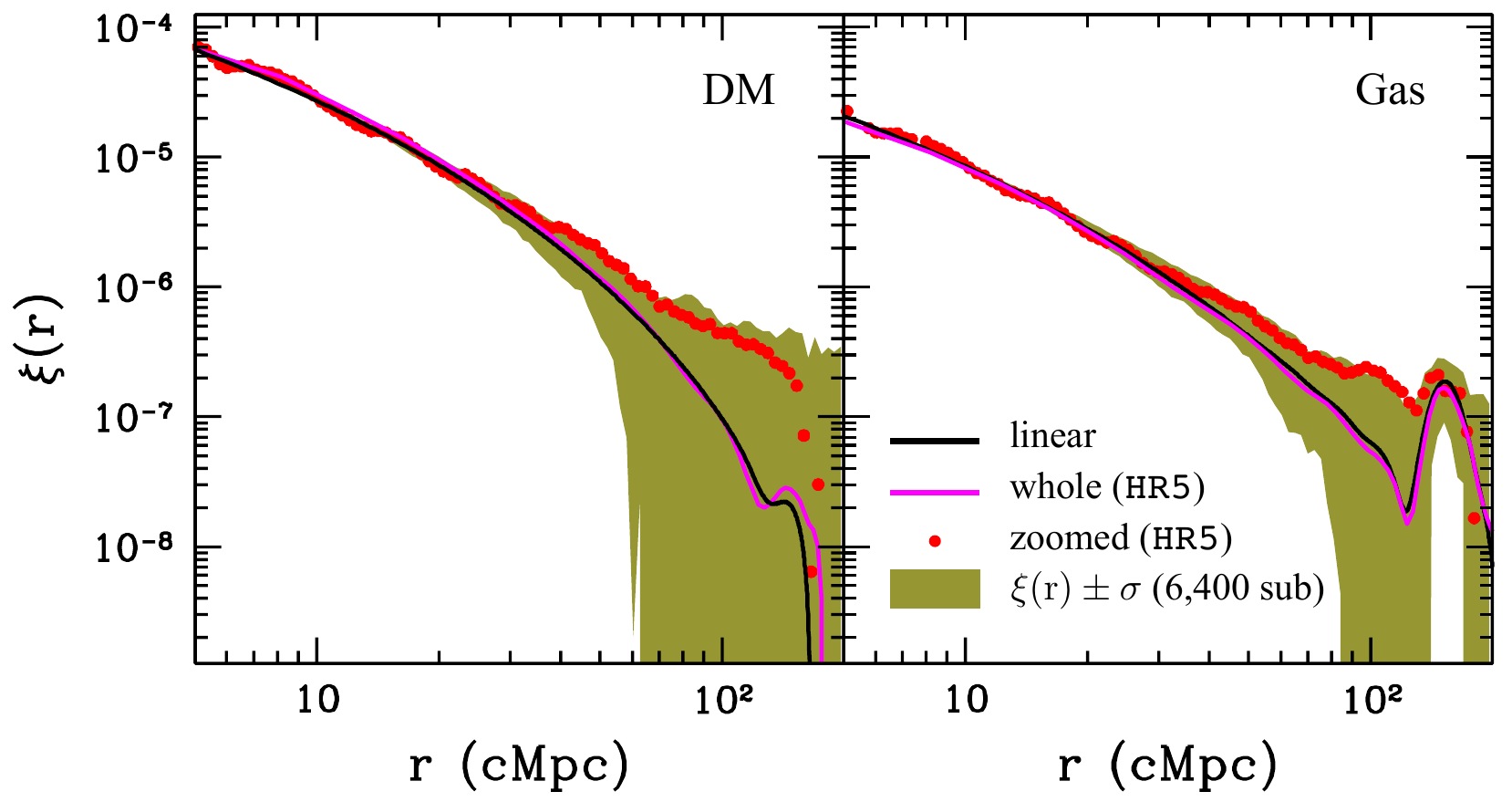}
    \caption{Cosmic variance presented in the two-point correlations measured from the gas (right) and dark matter (left) density fields at $z=200$. The olive shades show the $1\sigma$ distribution of the two-point correlations in the 6,400 sub-volumes sampled from 100 ICs generated using different random seeds and the magenta curves indicate those in the whole volume of \texttt{HR5} derived using the FFT. The black solid lines come from the linear theory and the red circles mark the measured correlations in the zoomed region of \texttt{HR5}. The two-point correlations in the zoomed region are situated within the $1\,\sigma$ scatter of the correlations of the sub-volumes.
    } \label{fig:cosvar}
\end{figure}

\section{Cosmic Variance of the Zoomed Region}\label{sec:cosvar}
In this section, we investigate the finite-volume effect of the zoomed region on the two-point correlations of the DM and gas density fields. In order to estimate the cosmic variance in the volume scale of the zoomed region, we measure the two-point correlations in 6,400 sub-volumes drawn from 100 ICs that have dimensions the same with that of \texttt{HR5} but are generated using varying random seeds. Each volume is uniformly binned into sixty four ($8\times8$) sub-volumes with a geometry similar with the zoomed region. The sub-volumes have the dimension of (262\,cMpc)$^3$ for each, being comparable to the volume of the zoomed region (240\,cMpc)$^3$. We constructed the density fields of DM and gas in $256^3$ grids for the whole volume and measured the two-point correlations of the density fields in each sub-volume. Figure~\ref{fig:cosvar} presents the two-point correlations from the linear model (black solid line), those from the whole volume of \texttt{HR5} (magenta curve), those measured in the zoomed region (red dots), and the $1\sigma$ scatter of the two-point correlations measured in the 6,400 sub-volumes (olive shades). One can see that the two-point correlations in the zoomed region (red dots) are situated inside the $1\sigma$ ranges. Therefore, even though the simulated correlations are substantially different to the linear theory in the zoomed region, the cosmic variance can account for the difference, within $1\sigma$ scatter.

\end{document}